\documentclass[twocolumn,aps,showpacs,amsmath,amssymb,prb]{revtex4}
\usepackage{graphics} 
\usepackage{psfrag}
\usepackage{epsfig}
\begin{document}
\date{\today}
\title{Enhancement of thermal transport in the degenerate periodic
                Anderson model}
\author{V. Zlati\'c$^{1,2}$, R. Monnier$^3$, J. K. Freericks$^4$}
\affiliation{$^1$Institute of Physics, Bijeni\v cka c. 46, 10001 Zagreb,
Croatia (permanent address), and}
\affiliation{$^2$International School for Advanced Studies (SISSA),
Via Beirut 2-4, I-34014 Trieste, Italy}
\affiliation{$^3$ETH H\"onggerberg, Laboratorium f\"ur
Festk\"orperphysik, 8093 Z\"urich, Switzerland}
\affiliation{$^4$Georgetown University, Washington D.C., USA}

\begin{abstract}
The low-temperature transport coefficients of the degenerate periodic $SU(N)$ 
Anderson model are calculated in the limit of infinite correlation between 
{\it f} electrons, within the framework of dynamical mean-field theory. 
We establish the Fermi liquid (FL) laws in the clean limit, 
taking into account the quasiparticle damping. 
The latter yields a reduced value of the Lorenz number in the 
Wiedemann-Franz law. Our results indicate that the renormalization of 
the thermal conductivity and of the Seebeck coefficient can lead to 
a substantial enhancement of the electronic thermoelectric figure-of-merit 
at low temperature.

Using the FL laws we discuss the low-temperature anomalies 
that show up in the electrical resistance of the intermetallic 
compounds with Cerium and Ytterbium ions, when studied as 
a function of pressure.  
Our calculations explain the sharp maximum of the coefficient 
of the $T^2$-term of the electrical resistance and the rapid variation 
of residual resistance found in a number of Ce and Yb intermetallics 
at some critical pressure.  

\end{abstract}

\maketitle

\section{Introduction 
\label{Sec: introduction}}
The low-temperature charge and thermal transport of heavy 
fermions and valence fluctuators with Ce, Eu, Yb and U ions 
display interesting and complex behavior, like a striking 
correlation~\cite{behnia.04,sakurai.05}  between the 
low-temperature Seebeck coefficient $\alpha(T)$ and the 
specific heat coefficient $\gamma=C_V/T$. 
In many of these systems the dimensionless ratio 
$q=|e|\lim_{T\to 0} \alpha/\gamma T$ is nearly the same, 
although the absolute values of  $\gamma$ and $\alpha/T$ 
vary by orders of magnitude. In metallic systems, the data 
show~\cite{hossain.04,sakurai.02,ocko.01,ocko.04,kohler.2008} 
small deviations from this universal behavior, due to the 
variation in carrier concentration, and in bad metals, $q$ 
can become quite large~\cite{behnia.04,sakurai.05}.
The Kadowaki-Woods (KW) ratio~\cite{kadowaki.87} which is 
defined as $\rho(T)/(\gamma T)^2$, where $\rho(T)$ is the 
electrical resistivity, exhibits similar universal 
features, if one takes into account the
effective low-temperature degeneracy of the {\it f} multiplet, 
as defined by the multiplicity of the crystal field (CF) ground 
state~\cite{kontani.04,KW-experiment}, the 
 carrier concentration and the unit cell volume~\cite{hussey.2005}. 

The near constancy of the KW and {\it q} ratios 
brings to the fore the validity of 
the Wiedemann-Franz (WF) law, $\kappa\rho / T ={\mathcal L}_0$, 
and a possibility of enhancing the electronic thermoelectric 
figure-of-merit in strongly correlated materials, 
$ZT=\alpha^2 T/\kappa\rho$, where $\kappa$ is the electronic 
contribution to the thermal conductivity and 
${\mathcal L}_0={\pi^2 k_B^2}/{3e^2}$ the Fermi liquid (FL) 
Lorenz number.
When the WF law holds, metals must have a thermopower larger than
155 $\mu$V/K to achieve $ZT>1$; to date no metal has been found with
so large a thermopower. In the temperature window where
the effective Lorenz number (${\mathcal L}=\kappa\rho/T$) is reduced,
one can
achieve $ZT>1$ with substantially lower thermopowers,
which might make it possible to find strongly correlated metals
that can be used for cooling applications at low temperature.

The above-mentioned universality of the KW and {\it q} ratios, 
typical of a Fermi liquid (FL)  state, attracted considerable 
theoretical attention.
Kontani~\cite{kontani.04} explained the KW ratio using
the orbitally degenerate periodic Anderson model that
he solved by the quasiparticle (QP) approximation of
Yosida and Yamada\cite{yamada.86,yamada.04}.
Neglecting both the vertex corrections and the momentum dependence
of the self-energy, he derived a KW ratio that depends on 
the degeneracy of the {\it f} states and brings the experimental 
data closer to the universal (theoretical) curve~\cite{KW-experiment}.
Miyake and Kohno~\cite{miyake.05} calculated the $q$ ratio 
for the same QP dispersion as in Ref.~\onlinecite{yamada.86},  
using an effective {\it N}-fold degenerate free-electron model 
in which the on-site correlation $U_{ff}$ is accounted for by 
the renormalized hybridization between the {\it c} and {\it f} 
states. Restricting the average number of {\it f} electrons 
to $n_f \leq  1$, they treated the QPs as free fermions and 
assumed that the repeated impurity scattering gives rise 
to an energy dependent relaxation rate.  
Neglecting the QP damping due to Coulomb repulsion, 
Miyake and Kohno~\cite{miyake.05} calculated the 
low-temperature thermopower as a logarithmic derivative 
of the frequency-dependent conductivity\cite{mott-jones} 
and show that the $q$ ratio is quasi-universal number. 

The effect of electron-electron scattering on the transport 
coefficients has recently been studied by Grenzebach et 
al.\cite{grenzebach.06},  using the dynamical mean-field 
theory~\cite{georges-kotliar-krauth-rozenberg.1996} (DMFT) 
of the periodic spin-1/2 Anderson model. 
The auxiliary impurity problem generated by the DMFT was 
solved by the numerical renormalization group (NRG) method, 
which discretizes the energy spectrum and defines the 
temperature as the difference between the two lowest energy 
states. This provides accurate results for the static 
properties at arbitrary temperature but cannot provide 
the thermal transport in the FL regime. The  DMFT+NRG 
calculations indicate an enhancement of the figure of 
merit when the temperature is reduced below its value at
the resistivity maximum, and show a 
breakdown of the WF law due to electron correlations.
Recently, we calculated the thermopower and the $q$--ratio 
of the periodic Anderson model taking into account 
the CF splitting\cite{zlatic.07b} and enforcing $n_f \leq 1$ 
at each lattice site (a realistic description of the CF states 
requires a local constraint). 
Solving the auxiliary impurity problem in the non-crossing 
approximation (NCA) we obtained a semi-quantitative description 
of the experimental data on heavy fermions and valence 
fluctuators in the incoherent regime. 
However, a detailed analysis of the DMT+NRG and the NCA 
results shows that nether method can describe the thermal 
transport much below the Kondo temperature $T_K$ nor 
establish the FL laws. 

In this paper, we discuss the coherent thermal transport of the 
periodic Anderson model with $SU(N)$ symmetry in the limit of 
an infinitely large Coulomb repulsion between $f$ electrons.
Such an effective $N$-fold degenerate model applies to
intermetallic compounds with Ce, Eu, Yb and U ions in the FL 
regime, where the excited CF states can be neglected~\cite{hewson}.
For a given compound, the number of effective  channels 
depends on the pressure and doping. We assume that the model 
has been solved in thermal equilibrium, so that the FL scale  
$T_0\sim 1/ \gamma$ is known. 
The total particle number per unit cell $n(\mu)$ is also $n$ known;  
$\mu$ denotes the chemical potential.
Alternatively, we can assume that the values of $\gamma$ and 
$n$ are taken from experiment. 
Using these equilibrium quantities, and enforcing the  constraint 
$n_f\leq 1$ at each lattice site,  we construct an analytic solution 
for the stationary heat and charge transport in the FL regime.

Unlike in previous work, we do not calculate the transport properties
in the QP representation, because the operator algebra in the
projected Hilbert space is not fermionic and the representation
of the exact charge and heat current density operators is cumbersome.
In the original fermionic representation, Mahan~\cite{mahan.98} has 
demonstrated that the correlation functions between the above current 
density operators  can be expressed in terms of transport integrals 
which differ only by powers of the excitation energy in their integrand,  
so that the same techniques can be used for their evaluation 
as in weakly correlated systems.

The transport coefficients are obtained by the DMFT and 
expressed in terms of the average conduction electron velocity, 
the renormalized density of conduction states, and the frequency 
and temperature dependent relaxation rate which explicitly 
takes into account the QP damping. Since we are considering 
the FL regime, we use the QP approximation of Yosida and 
Yamada\cite{yamada.86} to relate all these quantities to 
the FL scale $T_0$ and to the unrenormalized density of 
{\it c} states, ${\cal N}_c^0(\omega)$, evaluated at the shifted 
chemical potential $\mu_L=\mu+\Delta\mu$. 
The shift $\Delta\mu$ is determined by the Luttinger theorem. 
Using the Sommerfeld expansion, we obtain the universal FL laws 
which show that the KW ratio depends not only on the 
multiplicity of the $f$ state and the average FS velocity, 
but also on the carrier concentration and the unit cell 
volume, as observed experimentally~\cite{hussey.2005}. 
As regards the $q$-ratio, we find that changes in the carrier 
concentration, induced by pressure or chemical pressure, 
lead to deviations from universality.  
We also find large deviations from the WF law due to the lowering 
of the effective Lorenz number, which can lead to a substantial 
enhancement of the thermoelectric figure-of-merit $ ZT>1$.
The change in the effective degeneracy of the model induced 
by pressure explains the pronounced maximum of the coefficient 
of the $T^2$ term of the electrical 
resistance~\cite{jaccard.98,holmes.04,wilhelm.02,Wilhelm.05} 
and the rapid variation of residual resistance~\cite{holmes.04,jaccard.98}, 
found in a number of Ce and Yb intermetallics at some critical pressure.

The usefulness of the approximate analytic solution of the DMFT
equations becomes apparent when we realize that a numerical
calculation of the above mentioned universal ratios encounters
serious difficulty, and neither the NRG, nor exact diagonalization, 
nor quantum Monte Carlo approaches provide accurate transport 
coefficients in the FL regime (especially when the coherence 
temperature is low).
Combining analytical and numerical results enables a reliable
estimate of transport coefficients at arbitrary temperatures, which
is needed if the model is to be compared with experimental data.

The rest of this paper is organized as follows: 
Sec.~\ref{Sec: formalism} describes the DMFT calculations 
of transport coefficients in the low-temperature limit, 
in Sec.~\ref{Sec: FL_approach} we use the QP approximation 
to calculate the renormalized density of states and 
transport relaxation time in the FL regime, 
Sec.~\ref{Sec: FL_laws} establishes the FL laws, 
in Sec.~\ref{Sec: experimets} we use our results to discuss 
the experimental data, and Sec.~\ref{Sec: summary} 
provides the summary and conclusions.

\section{Formalism for the transport coefficients
\label{Sec: formalism}}
The $SU({{N}})$-symmetric periodic Anderson model is written in the standard
form~\cite{hewson},
\begin{equation}
                                \label{Hamiltonian}
\mathcal{H}
=
\mathcal{H}_{band}+\mathcal{H}_{imp}
+
\mathcal{H}_{mix}
-
\mu\mathcal{N} .
\end{equation}
Here, $\mathcal{H}_{band}$ describes the conduction ({\it c}) band with
unperturbed dispersion $\epsilon_{\bf k}$, assumed to be a even 
function of ${\bf k}$. 
Because we have an $SU(N)$ symmetric model, there are
$N$ distinct flavors of conduction electrons,
which are labeled by index $\sigma$, so that
\begin{equation}
 \mathcal{H}_{band}
=\sum_{\sigma=1}^N\sum_{\bf k}\epsilon_{\bf k}
c^\dagger_{\bf k\sigma}c^{}_{\bf k\sigma}.
\label{eq: ham_band}
\end{equation}
The operators $c^\dagger_{\bf k\sigma}$ ($c^{}_{\bf k\sigma}$)
create (annihilate) a band electron with momentum {\bf k}
and flavor $\sigma$. 
The non-interacting density of
conduction states ${\cal N}_c^0(\omega)$ is calculated for 
each flavor and any $\epsilon_{\bf k}$ as
$\mathcal{N}^0_c(\omega)
=
{1}/{{(\cal N}_i {\cal V})}
\sum_{\bf k}\delta(\omega-\epsilon_{\bf k})$,
where ${\cal N}_i$ is the number of lattice sites
and ${\cal V}$ the volume of the primitive unit cell.
The characteristic bandwidth of the unperturbed $c$-DOS 
is $W$. The number of conduction electrons is $Nn_c$ 
(per unit cell), i.e., there are $n_c$ conduction  
electrons of each flavor. 
We assume an infinite Coulomb repulsion, $U_{ff}\to \infty$,   
and  describe the localized  4{\it f}  states by
$\mathcal{H}_{imp}$,
\begin{equation}
 \mathcal{H}_{imp}
 =
\sum_{\sigma=1}^N \sum_i E_f^\sigma
\mathcal{P}f^\dagger_{i\sigma}f^{}_{i\sigma}\mathcal{P},
\label{eq: ham_imp}
\end{equation}
where $E_f^\sigma$ labels the site energy of flavor $\sigma$  
and $\mathcal{P}$ projects onto the subspace with zero or one 
total $f$-electrons per site. The number of  {\it f} electrons 
is restricted to $n_f\leq 1$, i.e., there are $n_f/N$ localized 
electrons per flavor. 
The hybridization between same flavor $c$ and $f$ electrons
is described by $\mathcal{H}_{mix}$
\begin{equation}
 \mathcal{H}_{mix}
=
\sum_{\sigma=1}^N\sum_{i,j}
V_{ij}^\sigma
\mathcal{P}
\left ( f^\dagger_{i\sigma}c^{}_{i\sigma}+
c^\dagger_{i\sigma}f^{}_{i\sigma}\right )
\mathcal{P},
\end{equation}
where the conduction-electron operators are now written 
in real space. The total electron number operator is 
$\mathcal{N}
=\sum_\sigma\sum_i(c^\dagger_{i\sigma}c^{}_{i\sigma}
+
f^\dagger_{i\sigma}f^{}_{i\sigma})$ and the chemical potential 
$\mu$ is adjusted to keep the total particle number 
$n=Nn_c+n_f$ constant (as a function of temperature or pressure). 
For a degenerate  paramagnetic state, all flavors are 
equivalent and the label $\sigma$ can be dropped.

The intermetallic compounds with 4$f$ ions are described by 
the $N$-fold degenerate model with on-site hybridization, 
$V_{ij}=V\delta_{ij}$, which simplifies the calculations. 
For large $N$, an exact solution is provided by the mean-field 
or the slave boson solution\cite{hewson} but for small $N$ the 
correlations give rise to the Kondo effect which cannot be described 
by mean field theories. The physically relevant value of $N$ 
depends on the effective degeneracy of the 4$f$ state. 
A single electron (or hole) in the lowest spin-orbit state 
of the 4$f$ shell of Ce (Yb) is $2J+1$-fold degenerate, 
where $J=5/2$ for Ce and $J=7/2$ for Yb. This degeneracy is 
further reduced by CF splitting and for small hybridization,  
and temperatures at which the excited CF states are unoccupied, 
the value of $N$ is determined by the degeneracy of the lowest CF state. 
If the hybridization is such that the CF splitting does not occur, 
or the excited states are thermally occupied, the effective degeneracy 
is defined by the lowest spin-orbit multiplet, which is a sextet 
for Ce and an octet for Yb. In all the physically relevant cases, 
$2\leq N\leq 8$, the system exhibits the Kondo effect. 

At high temperatures, the qualitative solution can be obtained 
by perturbative scaling\cite{hewson} 
which shows that the properties depend on an exponentially small 
energy scale $T_K\simeq W \exp(-1/Ng)$, where $g$ is the 
dimensionless coupling constant $g = V^2{\cal N}_c^0(\mu)/E_f$. 
At temperature $T_K$ the perturbation theory breaks down.  
The low-temperature thermodynamic quantities of the $SU(N)$ model 
are given very accurately by the DMFT+NRG and we assume that 
we know the numerical value of Fermi liquid scale 
$T_0={\pi^2k_B^2}/{3\gamma\cal{V}}$, where the linear coefficient of 
the low-temperature specific heat, $\gamma$, is measured per unit volume. 
Alternatively, $T_0$ can be estimated from the slave boson solution 
or taken from experiment. 
Unlike the situation for the static thermodynamic quantities,  
no currently available numerical method provides a reliable solution 
for the transport coefficients at low temperatures. 
In this paper, we evaluate the heat and charge transport of the $SU(N)$ 
symmetric Anderson model by the Fermi liquid theory\cite{yamada.86} 
which holds for $T\leq T_0$.  

For the static and uniform transport we need the ${\bf q}\to 0 $ 
component of the charge and heat current density operators 
which are obtained by commuting the Hamiltonian with the 
appropriate polarization operators\cite{mahan.81}. 
The total current density obtained in such a way satisfies
\begin{equation}
                                \label{current}
{\bf j}_c
=
\lim_{\tau,\tau^\prime\to 0}
\frac{eN}{{\cal V}{\cal N}_i}
\sum_{\bf k}{\bf v}_{\bf  k}
c^\dagger_{{\bf k}\sigma}(\tau) c^{}_{{\bf k}\sigma}(\tau^\prime) ,
\end{equation} 
where  ${\bf v}_{\bf  k}=\nabla\epsilon_{\bf k}/\hbar$
is the band velocity of the noninteracting
(and unhybridized) {\it c}-states.
For constant hybridization the energy current density is 
given by a similar expression which shows that the model 
satisfies the Jonson-Mahan theorem\cite{mahan.98}. 
This allows us to express the charge conductivity by 
$\sigma(T)= e^2 N L_{11}$,
the thermopower by 
$\alpha(T) |e| T = -  L_{12}/ L_{11}$,
and the electronic contribution to the thermal conductivity by
$\kappa(T) T = N ( L_{22}   -  L_{12}^2/ L_{11} )$.
In each of these expressions we have introduced
the (single-flavor) transport integrals:
\begin{equation}
                             \label{mj}
L_{mn}
=
\int d\omega
                  \left(-\frac{df}{d\omega}\right)
                   {\omega}^{m+n-2} \Lambda(\omega,T) .
\end{equation}
where $f(\omega)=1/[1+\exp(\beta\omega)]$ is the Fermi-Dirac
distribution function, $\omega$ is measured with respect
to the chemical potential $\mu$, and $\Lambda(\omega,T)$ 
is a function calculated in Appendix~\ref{correlation_function} 
by the Kubo linear response theory. 

At low temperature $(-df/d\omega)$ approaches a delta function 
and the main contribution to the integrals in Eq.~\eqref{mj} 
 comes from the low-energy excitations within the 
Fermi window, $|\omega|\lesssim k_BT$, and with the wave vectors 
in the vicinity of the Fermi surface. 
In the Fermi liquid state $\Lambda(\omega,T)$ can 
be calculated in a straightforward  way (for details  see 
Appendix~\ref{correlation_function}) 
which yields in the $\omega,T\rightarrow 0$ limit 
\begin{equation}
                             \label{lambda-FL}
{\Lambda(\omega,T)}
=
\frac{1}{3}{v^2_F}  {\cal N}_c(\omega)\tau(\omega,T),    
\end {equation}
for a three-dimensional system.
Here,  ${v^2_F} $ denotes the average of  $v_{\bf k}^2$ over 
the renormalized Fermi surface of hybridized states,
$\tau(\omega,T)$ is the transport relaxation time, given by the 
momentum-independent self-energy of the {\it c} electrons, 
\begin{equation}
                             \label{tau}
\tau(\omega,T)
=
\frac {-\hbar} { {\rm Im} \ \Sigma_c(\omega^+,T) } , 
\end{equation}
and ${\cal N}_c(\omega)$ is the renormalized DOS of $c$ 
electrons (for a single flavor),  
\begin{equation}
                             \label{eq: N_condution}
{\cal N}_c(\omega)
=
\frac{1}{{\cal V}{\cal N}_i}
\sum_{\bf k}  A_c({\bf k},\omega) . 
\end{equation}
The spectral function of the {\it c} electrons 
\begin{equation}
A_c({\bf k},\omega)
=
-\frac 1 \pi {\rm Im}\ G_c({\bf k},\omega^+)
\end{equation}
is obtained by evaluating the retarded Green's function 
$G_c({\bf k},z)$ just above the real axis, 
where $z=\omega^+=\omega+i\delta$ and $\delta\rightarrow 0$ 
from positive values.

The Green's functions of the periodic Anderson model are 
defined by the time-ordered thermal averages in the usual way  
and are cast in the Dyson form using the equation of motion 
or the diagrammatic expansion. 
The Dyson equations for {\it c} and {\it f} 
Green's function read~\cite{yamada.86,yamada.04} 
\begin{equation}
                             \label{G_c}
G_c({\bf k},z)
  =
\frac{z - E_f -\Sigma_f(z)+\mu}
{[z -\epsilon_{\bf k}+\mu]
[z -E_f -\Sigma_f(z)+\mu]-V^2} ,
\end {equation}
and
\begin{equation}
                             \label{G_f}
G_f({\bf k},z)
=
\frac{z -\epsilon_{\bf k}+\mu}
{[z -\epsilon_{\bf k}+\mu]
[z - E_f -\Sigma_f(z)+\mu]-V^2},
\end{equation}
where  $z$ denotes a variable in the complex energy plane 
and $\Sigma_f(z)$ the local self-energy of $f$ electrons 
which describes the renormalization of the (hybridized) 
$f$ states due to local Coulomb interaction. 
The Dyson equation for the conduction electrons
can also be written in the alternative form 
\begin{equation}
                             \label{G_c_standard}
  G_c({\bf k},z)
  =
\frac{1}
{z -\epsilon_{\bf k}+\mu-\Sigma_c(z)},
\end {equation}
where the local self-energy of the conduction electrons satisfies
\begin{equation}
                             \label{eq: sigma_c}
\Sigma_c(z) =\frac{V^2}{z-E_f  + \mu -\Sigma_f(z)}.
\end {equation}
This self-energy describes the renormalization of the unhybridized 
$c$ states due to the scattering on the $f$ states, i.e., 
$\Sigma_c$ includes the hybridization {\it and} the correlation effects. 
The DMFT condition ensures $\Sigma_f(\omega^+,T)$ to be local 
but the locality of $\Sigma_c(\omega^+,T)$ also requires the 
hybridization to be momentum independent. 
The dc transport relaxation time is obtained
by substituting $\Sigma_c(\omega^+,T)$
into Eq.~\eqref{tau} which yields
\begin{equation}
                             \label{tau_delta}
\frac {\tau(\omega,T)}{\hbar}
\simeq
\frac{(\omega-E_f - {\mathrm Re}\Sigma_f+\mu)^2
+(\delta-{\mathrm Im}\Sigma_f)^2}
{ V^2 (\delta-{\mathrm Im} \ \Sigma_f ) } , 
\end {equation}
where the $T,\omega$-dependence is due to $ \Sigma_f(\omega,T)$.
To account for the residual impurity scattering, 
which is present in any sample, we assume that $\delta$ in 
Eq.~\eqref{tau_delta} has a small but finite value. 
For temperatures such that 
$|{\mathrm Im} \ \Sigma_f(\omega,T)| \gg \delta$ this additional 
scattering is neglected and we discuss the temperature dependence 
of transport coefficients by setting $\delta=0$. 
At T=0 the self energy vanishes and $\tau(\omega,T)$ becomes a constant.   
The pressure dependence of the residual resistance is discussed 
using Eq.~\eqref{lambda-FL} with a constant relaxation time, 
which neglects the renormalization effects due to the repeated 
scattering on the impurities.  Where appropriate, the clean limit
implies we are in the regime where $\delta\ll |{\mathrm Im} \ \Sigma_f(\omega,T)|$,
whereas, we must examine the dirty limit, where $\delta > |{\mathrm Im} \ \Sigma_f(\omega,T)|$,
to describe residual resitance data.
\section{The Fermi liquid approach
\label{Sec: FL_approach} }

In the FL regime, the imaginary part of $\Sigma_f$ is small 
and can be neglected when discussing the renormalized
excitation spectrum in the limit $T,\omega\to 0$. 
Expanding $\Sigma_f(\omega)$ through linear order in frequency 
we find in that limit 
\begin{equation}
 \omega- [E_f + {\mathrm Re} \ \Sigma_f(\omega) - \mu]
\approx
(\omega-\tilde\omega_f)Z_f^{-1}+\mathcal{O}(\omega^2),
                  \label{eq: renormalized_omega_f}
\end{equation}
with
$Z_f^{-1}=[1-\partial\Sigma_f/\partial\omega|_{\omega=0}]$
being the enhancement factor ($0\le Z_f\le 1$)
and $\tilde\omega_f= [E_f+{\rm Re}\Sigma_f(0)-\mu]Z_f$.
The parameters $\tilde\omega_f$ and $Z_f$ appear in the 
the description of the model at low-temperatures.  

If we ignore the imaginary part of the self-energy, 
both Green's functions in Eqs.~\eqref{G_c} and ~\eqref{G_f}
develop poles\cite{yamada.86} which define the QP excitations
with wave vector $\bf k$.
Using Eq.~\eqref{eq: renormalized_omega_f}
we can  write the secular equation 
for the QP excitations in terms of the renormalized 
quantities, as
\begin{equation}
                    \label{eq: qp_tilde_condition}
 (\omega-\epsilon_{\bf k}+\mu)(\omega-\tilde\omega_f)-\tilde V^2=0,
\end{equation}
with $\tilde V=V\sqrt{Z_f}$ the renormalized hybridization.
The roots $\omega=\Omega^{\pm}_{\bf k}$ describe two QP branches
\begin{equation}
 \Omega^{\pm}_{\bf k}=\frac{1}{2}
\left[
(\epsilon_{\bf k}-\mu +\tilde\omega_f)
\pm \sqrt{(\epsilon_{\bf k}-\mu-\tilde\omega_f )^2+4\tilde V^2}
\right] 
\label{QP-dispersion}
\end{equation}
separated by the hybridization gap 2$\tilde V$. 
These excitations are only defined for energies close to 
the chemical potential, where ${\mathrm Im}\ \Sigma_f$
can be neglected. For large $\omega$ the QP approximation
breaks down and Eq. \eqref{eq: qp_tilde_condition}
is unphysical.
At the FS we have $\Omega^-_{ {\bf k}_F} =0$ or $\Omega^+_{{\bf k}_F}=0$, 
depending on the value of $Nn_c+n_f$. 
For Ce and Eu systems we place the bare $f$ level 
below the chemical potential, which puts $\mu$ close to the 
top of the lower QP branch. From $\Omega^-_{{\bf k}_F}=0$ we see 
that ${{\bf k}_F}$ is then close to the zone boundary, 
where $\epsilon_{{\bf k}_F} -\mu \simeq W$ ($W$ denotes the 
half-width of the unperturbed $c$ band).   
For Yb compounds, the bare level is above $\mu$, so that 
the lower QP branch is full and the chemical potential is 
close to the bottom of the upper branch. 
The equation $\Omega^+_{{\bf k}_F}=0$ gives ${{\bf k}_F}$ 
close to the zone center, such that $\epsilon_{{\bf k}_F} -\mu\simeq -W$. 
Close to the FS, the QP dispersion is very weak and  
$\Omega^\pm_{\bf k}$ describes two heavy QP bands 
with a half-width equal to $\tilde V_f^2/W$. 

The above derivation treats the QP as a non-interacting Fermi gas 
with effective parameters $\tilde\omega_f$ and $Z_f$ (or $\tilde V$).
These parameters can be related to the linear coefficient of the 
specific heat, which is assumed to be known.  
Using the expression for total QP density of states 
\begin{equation}
                 \label{eq: QP_dos}
\mathcal{N}^{QP}(\omega)
=
\frac{1}{{\cal V}{\cal N}_i}\sum_{\bf k}\delta(\omega-\Omega^{\pm}_{\bf k})
\end{equation}
and the fact that the QP excitations defined by 
Eq.~\eqref{eq: qp_tilde_condition} are infinitely long-lived 
(${\mathrm Im}\ \Sigma_f(\omega)=0$ at $T=0$)  
we write 
\begin{equation}
                   \label{gamma_QP}
\gamma
=
\frac{\pi^2  k_B^2}{6} N  \mathcal{N}^{QP}(0)
=
\frac{\pi^2k_B^2}{3{\cal V}}\frac{1}{k_{B}T_0},
\end {equation}
where the $N$-fold degeneracy of the system
has been taken into account. The relationship 
between $\gamma$ and $\mathcal{N}^{QP}$ is derived 
assuming that thermally excitated QPs increase the 
average energy per unit volume by approximately
$\Delta E\simeq (k_B T)^{2} N \mathcal{N}^{QP}(0)$.
The FL parameter $T_0$ introduced in Eq.\eqref{gamma_QP} 
defines the low-temperature scaling behavior of all physical 
quantities. In many systems, it also sets the temperature 
at which the QP description breaks down. 
This temperature might or might not coincide with the 
high-temperature Kondo scale of the system. 
By definition, the inverse of $T_0$ is given by the specific 
heat coefficient or, equivalently, by the density of the QP 
excitations of the $SU(N)$ model, which can be calculated 
very accurately in thermal equilibrium. 
As shown below, the low-temperature thermal transport is 
also characterized by $T_0$. 

The FL scale $T_0$ can be related to the partial densities 
of $f$ and $c$ states\cite{yamada.86} by expanding 
Eqs.~\eqref{G_c} and ~\eqref{G_f} for small $\omega$.
This yields the spectral functions,
\begin{eqnarray}
 A_c({\bf k},\omega)
&\approx&
a^c_{\bf k}(\omega)\delta(\omega-\Omega^\pm_{\bf k})
\label{eq: gc_qp2}
\end{eqnarray}
and
\begin{eqnarray}
 A_f({\bf k},\omega)
&\approx&
a^{f}_{\bf k}(\omega)\delta(\omega-\Omega^\pm_{\bf k}),
\label{eq: gf_qp2} 
\end{eqnarray}
where we used the fact that $\omega =\Omega^\pm_{\bf k}$
can only be satisfied with one of the roots,
and introduced the coefficients
\begin{equation}
 a^c_{\bf k}(\omega)
=
\left[1+
\frac{\tilde V^2}{(\omega-\tilde\omega_f)^2}
\right ]^{-1}
\end{equation}
and
\begin{equation}
 a^{f}_{\bf k}(\omega)
=
Z_f\frac{\tilde V^2}{(\omega-\tilde\omega_f)^2}
a^{c}_{\bf k}(\omega).
\end{equation}
These coefficients satisfy the FL sum rule\cite{yamada.86},
\begin{equation}
                  \label{a-sum-rule}
 a^{c}_{\bf k}(\omega)+ Z_ f^{-1}a^{f}_{\bf k}(\omega)=1 .
\end{equation}
When we substitute Eq.~\eqref{a-sum-rule} into Eq.~\eqref{eq: QP_dos} 
and use Eq.~\eqref{gamma_QP}, the summation of $ A_c({\bf k},\omega)$ 
and $ A_f({\bf k},\omega)$ over the first Brillouin zone yields  
\begin{equation}
\frac{2}{\cal V}\frac{1}{Nk_{B} T_0}
=
\mathcal{N}_c(0)
+
Z_f^{-1}\mathcal{N}_f(0),
                                 \label{eq: T_0_vs_N_f} 
\end{equation}
which shows that the  FL scale is set by the product 
of the large enhancement factor $1/Z_f$
and $\mathcal{N}_f(0)$.

To relate $T_0$ to the renormalized parameters  
$\tilde\omega_f$ and $Z_f$, we express, first, 
the renormalized {\it c} and {\it f}  DOS in terms 
of the density of unhybridized band states.  
Using the identity 
$\delta(\omega-\Omega^\pm_{\bf k})
=
\delta({\epsilon_{\bf k}}-{\epsilon_{\bf k}}^\pm_\omega)
/|d\Omega^\pm_{\bf k}/d\epsilon_{\bf k}|$
and
$d\Omega^\pm_{\bf k}/d\epsilon_{\bf k}
= a^{c}_{\bf k}(\omega)|_{\omega=\Omega^\pm_{\bf k}}$,
where ${\epsilon_{\bf k}}^\pm_\omega$ is the solution of
Eq.~\eqref{eq: qp_tilde_condition} for a given
(small) $\omega$, we obtain from 
Eqs.~\eqref{eq: gc_qp2} and ~\eqref{eq: gf_qp2} 
the result
\begin{equation}
                   \label{eq:N_c_omega}
\mathcal{N}_c(\omega)
=
\mathcal{N}^0_c\left(
\omega+\mu-\frac{\tilde V^2}{\omega-\tilde\omega_f}
\right ),
\end{equation}
and 
\begin{equation}
                   \label{eq:N_f_omega}
\mathcal{N}_f(\omega)
=
\frac{Z_f\tilde V^2}{[\omega-\tilde \omega_f]^2}\mathcal{N}_c(\omega). 
\end{equation}
The expressions in Eqs. \eqref{eq:N_c_omega} and \eqref{eq:N_f_omega} 
are similar in  spirit and in form to those in 
Refs.~\onlinecite{logan.04} and ~\onlinecite{logan.05},
where the fully interacting DOS of the periodic Anderson
model is equated to the DOS of a non-interacting $U=0$
model. The Green's functions of that model are defined 
by Eqs.~\eqref{G_c} and ~\eqref{G_f}  for $\Sigma_f(\omega)=0$.
Here, we find it more convenient to relate 
$\mathcal{N}_c(\omega)$ to the DOS of an {\it unhybridized} 
conduction band which is obtained from Eq.~\eqref{G_c_standard} 
for $\Sigma_c(\omega)=0$.
At low temperatures, where only the hybridized bands have
physical relevance, this auxiliary conduction band is
just a convenient mathematical construct.
But at high temperatures, where the model has to account
for the scattering of conduction states on localized
paramagnetic {\it f} states, the unhybridized conduction
band \textit{is} physically relevant.

The auxiliary model provides the renormalized Fermi surface (FS) 
of the periodic Anderson model simply by the Luttinger theorem,
i.e., from  the fundamental Fermi liquid relation which states that 
the volume of the FS cannot be changed by interactions. 
Using the standard form of the {\it c} electron Green's function 
[in Eq.~\eqref{G_c_standard}] and the fact
that $\Sigma_c(z)$ is momentum independent (this holds
for momentum-independent hybridization in infinite dimensions) we immediately
learn that the FS of the auxiliary band model (with
$n=Nn_c+n_f$ electrons per unit cell) coincides with the FS of the
periodic Anderson model with the same number of electrons per cell.
The shape of the renormalized FS is obtained by solving
Eq.~(\ref{eq: qp_tilde_condition}) at $\omega=0$, which yields
the implicit equation 
\begin{equation}
                                      \label{FS}
\epsilon_{{\bf k}_F}=\mu + \frac{\tilde V^2}{\tilde\omega_f}
\end{equation}
in terms of the non-interacting dispersion. We recall that 
$\mu$ is fixed by the condition
\begin{equation}
                                      \label{n}
n=
Nn_c+n_f
=
N {\cal V}\int_{-\infty}^{\mu} d\omega \ 
[{\cal N}_c(\omega)+{\cal N}_f(\omega)].
\end{equation}
The auxiliary model has exactly the same FS,
determined by the equation
\begin{equation}
                                      \label{FS_0}
\epsilon_{{\bf k}_F }=\mu_L,
\end{equation}
where $\mu_L$ is obtained from the integral 
\begin{equation}
                                      \label{n_0}
n
=
N {\cal V}\int_{-\infty}^{\mu_L} d\omega \ {\cal N}_c^0(\omega), 
\end{equation}
i.e., $\mu_L$ is the chemical potential of a conduction band 
with $n=n_c+n_f$ electrons.  
The same chemical potential $\mu$ ($\mu_L$)
appears in Eqs.~\eqref{FS}  and \eqref{n} 
[Eqs.~\eqref{FS_0}  and \eqref{n_0}],
because Luttinger's theorem ensures that the number
of {\bf k}-points enclosed by the FS coincides 
with the total number of electrons in the system.
Equations ~\eqref{FS} and \eqref{FS_0} give the shift 
 \begin{equation}
                                      \label{Delta_mu}
\Delta\mu=\mu_L-\mu= \frac{\tilde V^2}{\tilde\omega_f}.
\end{equation}
Substituting Eqs.~\eqref{eq:N_c_omega} and ~\eqref{eq:N_f_omega} 
into Eq.~\eqref{eq: T_0_vs_N_f} and using Eq.~\eqref{Delta_mu} 
to eliminate ${\tilde V^2}/{\tilde\omega_f}$, 
we find the relationship between $\tilde\omega_f$ and $T_0$, 
\begin{equation}
\tilde\omega_f
=
\Delta\mu
\frac{N \mathcal{N}_c^0(\mu_L){\cal V}k_{B} T_{0}/2}
     {1-N\mathcal{N}_c^0(\mu_L){\cal V}k_{B}T_{0}/2}
\simeq
\frac{N}{2} \Delta\mu \mathcal{N}_c^0(\mu_L){\cal V} k_{B}T_0 , 
                                 \label{eq: tilde_omega_T0} 
\end{equation} 
where the last expression neglects the exponentially 
small term in the denominator. The sign of $\tilde\omega_f$ 
is set by  $\Delta\mu$ which is positive for Ce and Eu compounds 
and negative for Yb compounds. The enhancement factor is obtained 
from Eq.~\eqref{Delta_mu} as $Z_f= \Delta\mu \ \tilde\omega_f/V^2$. 

The ratio $2\tilde\omega_f/T_0=\Delta\mu {\cal V} \mathcal{N}_c^0(\mu_L)$ 
depends, for a given $n(\mu)$, on the renormalized chemical 
potential $\mu$, the auxiliary one $\mu_L$, 
and the unit cell volume ${\cal V}$. 
While $\mu_L$ is easily obtained 
from $n(\mu)$, the value of  $\mu$ is difficult 
to find without numerical calculations.
Except, for very small $\tilde V$, large $N$, and 
for  $n_f\simeq 1$ and $n_c\simeq 1/2$, when each 
conduction band is close to half-filling. 
In that case, the renormalization does not change much 
the partial occupancies of the {\it f}  and  {\it c} states, 
and we can approximate ${n_c} \simeq 
{\cal V}\int_{-\infty}^{\mu} d\omega \mathcal{N}_c^0(\omega)$, 
which yields 
\begin{equation}
                         \label{eq: n_f_approx}
N {\cal V} \int_{\mu}^{\mu_L} d\omega \mathcal{N}_c^0(\omega)
 \simeq
{n} - N{n_c}= {n_f} ,
\end{equation}
where we have implicitly assumed $\mu_L>\mu$, which applies
to Ce and Eu compounds. A similar result can be obtained for Yb
compounds after an electron-hole transformation.
For large $N$ the integral is small and since 
the integrand  has a maximum around $\mu$ 
(the auxiliary band is close to half-filling), 
it follows that $\Delta\mu$ is small. Using 
$\mathcal{N}_c^0(\mu_L) \simeq \mathcal{N}_c^0(\mu)$
we estimate 
\begin{equation}
                         \label{eq: large N}
{\cal V}\int_{\mu}^{\mu_L} d\omega \mathcal{N}_c^0(\omega)
\simeq
{\Delta\mu{\cal V}\mathcal{N}_c^0(\mu_L)}
\simeq 
\frac{n_f}{N} 
~. 
\end{equation}
Thus, for large $N$ and $n_f\simeq 1$, we have a simple 
relation $\tilde\omega_f\simeq n_f k_B T_0/2$. 
For small $N$, and/or small $n_f$, Eq.~\eqref{eq: n_f_approx} 
still holds but we cannot claim that $\Delta\mu$ is small.  
If $\mathcal{N}_c^0(\omega)$ decreases rapidly for 
$\omega > \mu$, as it usually does, we can conclude 
$\Delta\mu {\cal V} \mathcal{N}_c^0(\mu_L) \ll n_f/N 
\ll \Delta\mu {\cal V} \mathcal{N}_c^0(\mu)$ 
but cannot express 
$
{\Delta\mu{\cal V}\mathcal{N}_c^0(\mu_L)}$ 
in terms of $n_f$ or relate $\tilde\omega_f$ to $ T_0$ 
in a simple way, as in the large-$N$ limit. 

The auxiliary model with unhybridized conduction 
electrons has a simple physical interpretation 
for systems with small hybridization and $n_f\simeq 1$,  
i.e., for heavy fermions with very low Kondo temperature.  
In such systems the 4$f$ ions are in a well defined 
valence state, the low-energy dynamics is dominated by 
spin fluctuations, and $n_f$ is temperature-independent\cite{hewson}.
At low temperatures, the renormalized FS is defined by the set of 
${\bf k}$ vectors satisfying $\epsilon_{\bf k}=\mu+\Delta\mu$.
This FS is large, because it encloses $(n_c+n_f/N)$ states of each flavor. 
At high temperatures, the Fermi surface is small, because 
it encloses only $n_c$ conduction states ($f$ states are localized and 
do not contribute to the Fermi volume).  
Ignoring the fact that thermal fluctuations remove the discontinuity 
in the Fermi distribution function, we approximate the renormalized FS 
by the FS of unhybridized conduction states. (This holds for $T\ll W$.) 
On the other hand, the condition $\epsilon_{\bf k}=\mu$ 
defines a set of {\bf k} vectors which are close to the FS 
of the auxiliary band with $n_c$ electrons. This FS is also 
small, because it encloses only $n_c$ states of each flavor. 
Since $n_f$ and $n_c$ are temperature-independent, 
the shift $\Delta\mu$ provides the difference between 
the FS of the high-temperature paramagnetic phase  
with $Nn_c$ conduction electrons (and $n_f$ localized states), 
and the FS of the low-temperature FL phase with $N$ bands
containing each $n_c+n_f/N$ hybridized states. 
The above considerations show that the formation of the 
coherent QP bands is accompanied by a 'jump' of the Fermi volume 
from a small to a large value, and that the large FS encloses 
$n_f/N$ additional states. 
The auxiliary model approximates the conduction band of the 
periodic Anderson model and allows us to estimate this jump. 

So far, we neglected the imaginary part of self-energy, 
since we considered only the low-energy excitations at $T=0$ 
In order to calculate the transport properties at low 
but nonzero temperature, we have to include the QP damping 
which is given by the imaginary part of the self energy (in the clean limit). 
To estimate ${\mathrm Im} \ \Sigma_f(\omega,T)$ at low temperatures, 
we use the diagrammatic analysis of Yosida and Yamada\cite{yamada.86}.
In infinite dimensions, the expansion is in terms
of the local Green's functions, and produces
the FL expression~\cite{yamada.86,kontani.04}
\begin{equation}
                                  \label{Im_sigma}
-{\rm Im} \Sigma_f(\omega,T)
\simeq
\frac \pi 2  [\omega^2 + (\pi k_B T)^2]
(N-1)[{\cal V}{\cal N}_f(0)]^3\Gamma_f^2 ,
\end{equation}
where $\Gamma_f$ is the irreducible 4-point scattering vertex
for electrons with different flavors.
Eq.~\eqref{Im_sigma} is a straightforward generalization of
the result produced by second order perturbation theory in which
the bare interaction $U_{ff}$ is replaced by the scattering vertex 
$\Gamma_f$. 
In the limit of large correlations, when the charge fluctuations
are suppressed, the Ward identity~\cite{yamada.86,yamada.04}
gives $Z_f^{-1}=(N-1){\cal V}{\cal N}_f(0)\Gamma_f$, such that
$\gamma=(\pi^2k_B^2/6 ) N (N-1){\cal V}[{\cal N}_f(0)]^2 \Gamma_f$,
where we neglect the much smaller conduction electron
contribution to $\gamma$.
Substituting into Eq.~(\ref{Im_sigma}) yields in the $T,\omega\to 0$ 
limit\cite{kontani.04}
\begin{equation}
                    \label{eq: Im sigma_f}
 {\mathrm Im}\ \Sigma_f(\omega,T)
\simeq
-2 \pi \frac{\omega^2
+ (\pi k_B T)^2}{(N-1)N^2{\cal{V}}\mathcal{N}_{f}(0)(k_{B} T_0)^2}.
\end{equation}
Expanding $\Sigma_c(\omega,T)$ in Eq.~\eqref{eq: sigma_c} 
into a power series 
[with the real and imaginary part of $ \Sigma_f(\omega,T)$ given by 
Eqs.~ \eqref{eq: renormalized_omega_f} and  \eqref{eq: Im sigma_f}, 
respectively]   
we obtain the self energy of $c$ electrons in the FL form. 
That is, in the FL regime, the real part of $\Sigma_c(\omega,T)$ 
is a constant  and the imaginary part is a quadratic function 
of  $T$ and $\omega$. 

Inserting $\Sigma_f(\omega,T)$ in Eq.~\eqref{tau_delta} for 
$\tau(\omega,T)$ and using 
${Z_f\tilde V^2}/\mathcal{N}_f(0) ={\tilde \omega_f^2}/\mathcal{N}_c(0)$ 
yields for $T, \omega \to 0$ the leading term 
\begin{eqnarray}
                \label{eq: tau_low_T_omega}
\tau(\omega,T)
&=&
\frac{ \hbar (N-1)N^2{\cal{V}}\mathcal{N}_c^0(\mu_L)k_B^2 T_0^2}
{2\pi\tilde\omega_f^2} 
\frac{(\omega-\tilde\omega_f)^2}{(\pi k_BT)^2+\omega^2} 
\nonumber\\
&\simeq&
\tau_0(T)
(1-\frac{\omega}{\tilde\omega_f})^2
(1-\frac{\omega^2}{\pi^2 k_B^2 T^2}) , 
\end{eqnarray}
where 
\begin{equation}
                \label{eq: tau_00}
\tau_0(T)
=
\frac{ \hbar (N-1)N^2{\cal{V}}\mathcal{N}_c^0(\mu_L)}{2\pi^3}
\frac{T_0^2}{ T^2} .
\end{equation}
The second line in Eq.~\eqref{eq: tau_low_T_omega} emphasizes 
the fact that the limit $\omega\to 0$ is taken before $T\to 0$ 
and that $\Sigma_f(\omega^+,T)$ is only known up to the $\omega^2$ terms, 
so that the Sommerfeld expansion cannot be extended beyond the second 
order (up to that order both forms produce the same result).  
The first (second) bracket in Eq.~\eqref{eq: tau_low_T_omega} 
is due to the real (imaginary) part of $\Sigma_f(\omega^+,T)$. 
Unlike $\Sigma_c(\omega,T)$, the FL laws produced by the Sommerfeld 
expansion of transport coefficients  are affected not only by the 
$\omega$-dependence of ${\mathrm Im }\ \Sigma_f(\omega,T)$ but of 
${\mathrm Re}\ \Sigma_f(\omega,T)$ as well.

In order to make an estimate of the slope of the renormalized 
$f$ DOS, which is needed for thermal transport, we invoke the 
DMFT condition. 
This condition also provides a physical interpretation 
of the low-energy parameter $\tilde\omega_f$.  
In DMFT, we compute the local Green's function
from the local self-energy
\begin{eqnarray}
                  \label{eq: g_f_loc}
 G_f(z)
&=&
\sum_{\bf k}G_f({\bf k},z)\\
&=&
{\cal V}\int d\epsilon
\mathcal{N}_c^0(\epsilon)
\frac{1}{z-E_f+\mu-\Sigma_f(z)-\frac{V^2}{z-\epsilon+\mu}},
\nonumber
\end{eqnarray}
map it onto the Green's function of an effective 
single impurity Anderson model with a hybridization 
function $\Delta(z)$
\begin{equation}
 F(z)=\frac{1}{z-E_f+\mu-\Delta(z)-\Sigma_f(z)}, 
\label{eq: dmft_imp}
\end{equation}
and adjust $\Delta(z)$ to make $G_f(z)$ and $F(z)$ identical. 
The DMFT procedure works because the large-dimensional limit
guarantees that the functional relationship between the local
self-energy and the local Green's function for
the lattice is identical to the functional relationship between
the self-energy for the impurity and the impurity Green's function.
In the limit where $T=0$ and $\omega\rightarrow 0$, we approximate
$\Delta(\omega)=i\Delta_0$ with the constant $\Delta_0<0$ and 
write the DMFT condition at low frequencies as
\begin{equation}
\mathcal{N}_f(\omega)
=-\frac{1}{\pi}{\rm Im}\ F(\omega^+)
\simeq \frac{1}{\pi|\Delta_0|}
\frac{\tilde\Delta_f^2}{(\omega-\tilde\omega_f)^2+\tilde\Delta_f^2 },
                       \label{eq: dmft_dos_w}
\end{equation}
where $\tilde\Delta_f=\Delta_0Z_f$.
This Kondo-like form of $\mathcal{N}_f(\omega)$ holds only for 
$\omega\ll \tilde\omega_f$, just like the quasiparticle dispersion 
makes sense only for $\omega\ll T_0$. 
It cannot be extrapolated to higher frequencies, where the 
approximation $\Delta(\omega)=i\Delta_0$ does not hold. 
The approximate form given by expression Eq.~\eqref{eq: dmft_dos_w} 
has a maximum at $\tilde\omega_f$, where the exact {\it f} \  DOS 
has a gap;  the exact DMFT spectral function has a maximum between 
$\mu$ and $\tilde\omega_f$.
The width of effective Kondo resonance is $\tilde\Delta_f$. 
In the FL regime, $\omega\leq k_B T \ll \tilde\omega_f$,  
the transport coefficients depend on the characteristic energy 
scales defined by this effective Kondo resonance. 

The width $\tilde\Delta_f$, like $\tilde\omega_f$,  can be 
related to the FL scale $T_0$ of the periodic Anderson model. 
Equating the {\it f}-DOS and the effective single 
impurity DOS at $\omega=0$  yields
\begin{equation}
 \pi{\Delta\mu} {\cal V}\mathcal{N}_c^0(\mu_L)
=\frac{x}{1+x^2}
                         \label{eq: dmft_dos_w=0_x}
\end{equation}
where $x=\tilde\omega_f/\tilde\Delta_f$.
Solving for $x$ produces the result
\begin{equation}
\tilde\Delta_f
=
\frac{2\pi\Delta\mu{\cal V}\mathcal{N}_c^0(\mu_L)}
{1\pm\sqrt{1-({2\pi\Delta\mu{\cal V}\mathcal{N}_c^0(\mu_L)}^2)}}
\tilde\omega_f
                         \label{eq: dmft_dos_tilde_w}
\end{equation}
and we choose the negative sign.
Since ${\Delta\mu{\cal V}\mathcal{N}_c^0(\mu_L)}$ was shown
to be very small
[see the discussion around Eqs. \eqref{eq: n_f_approx} and \eqref{eq: large N}]
the root in the above expression can be expanded
to produce the lowest order result,
\begin{equation}
                         \label{eq: tilde_delta_small_nc}
\tilde\Delta_f
=
\frac{\tilde\omega_f}
{\pi\Delta\mu{\cal V}\mathcal{N}_c^0(\mu_L)}
\simeq
\frac{1}{2\pi} {Nk_{B} T_0}. 
\end{equation}
The above expression, which follows from the DMFT condition, 
ensures that that the initial slope of the renormalized $f$ DOS 
is very small, 
\begin{equation}
                         \label{eq: Nf_prime}
\left[\frac{\partial \mathcal{N}_f(\omega)}
{\partial\omega}\right]_{\omega=0} 
\simeq 
\frac{\tilde\omega_f}{\tilde\Delta_f}
\simeq 
\pi\Delta\mu {\cal V} \mathcal{N}_c^0(\mu_L) \ll 1.
\end{equation}

Finally, we remark that the FS average of the 
unrenormalized velocity squared, $v_{\bf k}^2 $, 
can be found from the integral
\begin{equation}
                                      \label{v^2}
{v^2_F} 
=
\langle v^2_{{\bf k}_F} \rangle
=
\int d^{d}{\bf k} \ \delta({\bf k} - {\bf k}_F) \ v^2_{\bf k}, 
\end{equation}
where the $\delta$--function restricts the integral to the renormalized FS.
In infinite dimensions $v^2_F$ is a constant for all fillings but
in lower dimensions the change of the Fermi volume with
pressure or temperature can modify ${v^2_F} $ and
affect thermal transport.
Numerical calculations for the 3-dimensional periodic
Anderson model with nearest-neighbor hopping on
a simple cubic lattice give
$v_F^2 = ({t a_l}/{\hbar})^2 v^2$ where $v^2\simeq 1.4$
for $n\simeq 1/2$ and  $v^2\ll 1$ for $n\simeq 1$.

\section{The Fermi liquid laws
\label{Sec: FL_laws}}
The transport coefficients of the periodic $SU(N)$ Anderson
model with infinite correlation between $f$ electrons 
are obtained in the FL regime by making the lowest order
Sommerfeld expansion of transport integrals, Eq.~\eqref{mj}, 
and expressing the integrand $\Lambda(\omega)$ in terms of the known 
expressions for $v^2_{k_F}$, ${\cal N}_c(\omega)$, 
and  $\tau(\omega,T)$. The algebra is straightforward   
(for details see Appendix \ref{Transport_coefficients}), 
and yields the transport coefficients as simple 
powers of reduced temperature $T/T_0$. 

The specific resistance of $N $ parallel channels 
obtained in such a way is
\begin{eqnarray}
                             \label{eq: rho_tt}
{\rho(T)}
&=&
\frac{9\pi^3{\cal V}} 
{\hbar{e^2 {v^2_F}  }
N(N-1)[N{\cal V}\mathcal{N}_c^0(\mu_L)]^2}
\left(\frac{ T}{T_0}\right)^2  
\end {eqnarray}
which holds for $N\geq 2$ and arbitrary ${\cal N}_c^0(\omega)$. 

The resistivity expression in Eq.\eqref{eq: rho_tt} 
neglects terms of the order of $({T}/{ T_0})^4$ 
and is only valid for ${T}\ll T_0$.
Even in that temperature range,  $\rho(T)$ deviates 
from the universal (KW) form, because the pre-factor of the 
$(T/T_0)^2$ term has an explicit parameter dependence.   
The value of these parameters depends o the Fermi volume 
of the system, which can be changed in several ways.
For example, additional impurity scattering or lattice 
expansion (negative pressure) can localize the $f$ states 
and exclude them from the Fermi volume. Another possibility 
is to increase the effective degeneracy of the $f$ states 
(by pressure or thermal population of the excited states), 
which changes the number of resonant channels and shifts  
the Fermi surface away closer to the zone center. 
The `jumps' in  the Fermi volume changes $\mu_L$ and ${v^2_F} $,  
which has a strong impact on the resistivity.
This feature can be used to explain the resistivity 
anomalies that accompany  the localization or 
delocalization of {\it f} electrons in heavy fermions, 
i.e., the breakdown or the formation of the QP bands. 
The rapid change of the coefficient of the $T^2$-term in the resistivity 
following the pressure- or doping-induced `jump'  
of the Fermi volume in Cerium compounds, is discussed 
in more detail below. 
On the other hand, if we tune the model parameters of the $SU(N)$ model 
in such a way that the Fermi volume is preserved 
(by keeping $N/n$ constant), the Luttinger theorem ensures 
that $\mu_L$ and ${v^2_F} $ do not change. In that case, 
the pre-factor of $(T/T_0)^2$ is constant but $\rho(T)$ changes 
due to variations in $T_0$ which can be exponentially fast. 

For large $N$, the approximation 
$N {\cal V}\mathcal{N}_c^0(\mu_L)\simeq {n_f}/{\Delta\mu}$  
yields the expression 
\begin{eqnarray}
\rho(T)
&\simeq&
                             \label{eq: rho_gama}
\frac { 81 (\Delta\mu/n_f)^2 {\cal V}^3 }
{ \hbar N(N-1)\pi k_B^2 e^2 {v^2_F} }
(\gamma T)^2  
\end {eqnarray}
which simplifies the discussion of heavy fermions 
in the Kondo limit, $n_f\simeq 1$. 
The Luttinger theorem ensures ${n_f}/{\Delta\mu}$ 
is constant even when the tuning of the external parameters 
gives rise to a charge transfer between the {\it f} 
and {\it c} states and thereby changes the renormalized 
chemical potential. 
As long as the tuning does not affect the degeneracy 
of the ground state, the pre-factor of the 
$(\gamma T)^2$-term of the resistivity is always the same. 

The Seebeck coefficient is obtained by using the Sommerfeld 
expansion for $L_{12}$ and $L_{11}$ which gives 
\begin{eqnarray}
{\alpha(T)}
&\simeq&
                                  \label{eq: alpha.12_3}
\frac{4 \pi^2 k_B}{|e| N \Delta\mu{\cal V}\mathcal{N}_c^0(\mu_L)} 
\frac{T}{T_0} .
\end {eqnarray}
The enhancement of $L_{12}$ is solely due to the real part 
of the self energy. The imaginary part corrects $L_{11}$, 
and gives a factor (3/2) enhancement of $\alpha(T)$ 
with respect to the $U=0$ case. 
This factor does not arise in mean field theory or in  
slave boson approximations based on quadratic Hamiltonians 
with renormalized parameters~\cite{miyake.05}. 
The above result corrects the expression used in our previous 
paper~\cite{zlatic.07b}, which was derived neglecting the 
energy dependence of the density of states. 
The term $\Delta\mu$, which does not occur in the 
low-temperature resistivity expression in Eq. \eqref{eq: rho_tt}, 
has an explicit parameter dependence, so that, 
strictly speaking, $\alpha(T)$ is not a universal 
function of $T/T_0$. 
In bad metals and systems with a low-carrier concentration, 
$\mu_L$ is close to the band edge, where $\mathcal{N}_c^0(\mu_L)$ 
could be very small, so that $\alpha(T)$ could be very large. 

The Seebeck coefficient of heavy fermions with $n_f\simeq 1$ 
and large $N$ assumes the simple form 
\begin{equation}
                             \label{alpha-tau}
{\alpha(T)}
=
\mp
\frac{4\pi^2 k_B}{n_f|e|} 
\frac{ T}{T_0} . 
\end {equation}
Since the doubly occupied {\it f} states are removed from 
the Hilbert space, the model is highly asymmetric, and the 
initial slope $\lim_{T\to 0} \alpha(T)/T$ never vanishes. 
As a matter of fact, the closer the system is to half-filling, 
the larger is the slope, $\alpha/T\propto 1/T_0$. This, however, does not 
necessarily imply a large thermopower, since the FL laws are only 
valid for $T\ll T_0$ and close to half-filling $T_0$ is 
exponentially  small. 

The FL result for the thermal conductivity in the clean limit reads 
\begin{equation}
                                       \label{kappa2}
\kappa(T)
=
T \sigma(T)
{\cal L}_0(T) , 
\end {equation}
where the usual Lorentz number, $ {\cal L}_0={\pi^2 k_B^2}/{3e^2}$,  
has been replaced by the effective one, 
\begin{equation}
                                       \label{Lorentz}
{\cal L}_0(T)=\bar{\cal L}_0
\left[1- \frac{32\pi^2}{ n_f^2} \left( \frac{T}{ T_0}\right)^2\right] ~,
\end {equation} 
and $\bar{\cal L}_0={\pi^2 k_B^2}/{2e^2}$.
This change is due to the imaginary part of the self energy 
and therefore not obtainable by mean-field or slave-boson calculations 
which neglect the QP damping. 
The $T\to 0$ limit yields the Wiedemann-Franz law, 
$\kappa(T)\propto \ T \sigma(T)$,  
but the correction given by the square bracket leads 
to deviations even at low temperatures. 
Since the factor multiplying the $T^2$ term is very large,  
we find a reduction of $\kappa(T)$ and substantial deviations 
from the WF law much below $T_0$.

Before closing this section, we summarize the procedure 
for calculating the transport coefficients of 
the periodic Anderson model in the FL regime. 
The model is specified by the hybridization $V$, 
the bare {\it f}-level position $E_f$, the degeneracy $N$, 
the center of the bare conduction band $E_0$,   
the unperturbed dispersion $\epsilon_{\bf k}$, 
and the restriction $n_f\leq 1$.
The unperturbed  density of conduction states 
$N_c^0(\omega)$ is easily obtained from $\epsilon_{\bf k}$. 
For a given total electron density $n$ (per cell)
we  use the DMFT+NRG or some simpler scheme, 
like the slave bosons, to 
find the renormalized chemical potential, 
the number of {\it f} electrons, and the 
low-temperature coefficient of the specific 
heat which sets the FL scale $T_0$. 
These static quantities can be calculated 
in thermal equilibrium to a very high accuracy. 
For constant hybridization, we can make 
separate DMFT+NRG runs for $n_c(\mu)$ and $n_f(\mu)$  
and do not have to calculate the spectral function 
which is not given very accurately (for all $\omega$) by the NRG. From 
$n=Nn_c+n_f$, we obtain $\mu$, $\mu_L$, and 
$\Delta\mu$, which provide the renormalized FS 
and the average ${v^2_F} $. From $T_0$, 
$\mu_L$, $\Delta\mu$ we obtain the 
Kondo scale $\tilde\omega_f$ and specify completely 
the low-energy behavior of $N_c(\omega)$ and $\tau(\omega)$.  
The FL laws follow at once. 
For heavy fermions with large $N$, $n_f\simeq 1$ and 
small $\tilde V$, the procedure simplifies considerably, 
since we can approximate  
$\Delta\mu {N {\cal V}\mathcal{N}_c^0(\mu_L)}\simeq {n_f}$, 
and obtain all the renormalized quantities using 
$\tilde\omega_f \simeq n_f k_B T_0/2$.

\section{Discussion of  experimental results
\label{Sec: experimets}}

\subsection{The universal ratios}
The FL laws in Eqs.~(\ref{eq: rho_tt}--\ref{kappa2}) 
describe coherent charge and heat transport in 
stoichiometric compounds, in a way analogous to the 
phase-shift expressions for dilute Kondo alloys~\cite{hewson}.
They explain the near-universal behavior of 
the KW ratio~\cite{kadowaki.87,KW-experiment}
reported for many heavy xfermions and valence fluctuators.
The ratio ${ \rho(T) } / {(\gamma T)^2}$ obtained 
from Eq.~\eqref{eq: rho_tt} exhibits an explicit 
dependence on the ground state degeneracy and the 
average velocity (squared), and an implicit dependence 
on the Fermi volume, i.e., on the carrier concentration. 
The ${{N}}$-dependence as well as the dependence on 
carrier concentration of the KW ratio were recently 
emphasized by Kontani and his 
collaborators~\cite{kontani.04,KW-experiment}, 
who obtained the power law $n^{-4/3}$ for the latter 
by using the free electron approximation for 
the average velocity and the density of states in 
the expression for the resistivity.
Our formulation leads to a similar $N$-dependence 
[see also Eq.~\eqref{eq: rho_gama}] but includes also 
the dependence on the average velocity, the renormalized 
Fermi volume and the carrier concentration, which are 
implicit functions of the degeneracy and have to be taken 
into account when discussing the pressure or doping 
experiments on strongly correlated electron systems
~\cite{hussey.2005}. 

As regards the thermal transport, 
Eq.~\eqref{eq: alpha.12_3} gives 
$q=\lim_{T\to 0}|e\alpha|/\gamma T
\simeq 12/[N\Delta\mu\mathcal{N}_c^0(\mu_L)]$, 
which has an explicit parameter dependence, 
so that the $q$-ratio, like the KW ratio, 
can deviate from the universal value~\cite{behnia.04,sakurai.05}.
In bad metals and systems with a low-carrier concentration, 
$\mu_L$ is close to the band edge, where $\mathcal{N}_c^0(\mu_L)$ 
could be very small, making $\alpha(T)$ large. 
For a given system, the $q$-ratio can be pressure dependent 
due to the transfer of $f$ electrons into conduction band. 
The data on the pressure dependence are not available but 
the deviations from the universal value are indicated 
by recent chemical pressure data~\cite{sakurai.05}. 

A further remarkable consequence of correlations is 
the enhancement of the low-temperature figure-of-merit 
due to the deviations from the WF law.
Using Eq.~\eqref{kappa2} and neglecting phonons 
in the FL regime we express the figure-of-merit as a ratio 
$ZT= \alpha^2(T)/ {\cal L}(T) $. 
For constant Lorenz number, the maximum of $ZT$ is defined 
by the thermopower but in correlated systems the temperature 
dependence of the effective Lorenz number can lead to 
an additional enhancement.  
Even though our FL result is valid only for $T\ll T_0$, 
it captures the essential features: an increase of $\alpha^2$ and 
a decrease of ${\cal L}(T) $ that ultimately give rise to $ZT>1$.
The enhancement of $ZT$ is due to the renormalization of 
the thermopower {\it and} the Lorentz number 
i.e., $ZT > 1$ is not restricted to metallic systems with 
$\alpha(T) > 155 \ \mu$V/K.
We expect large $ZT$ for small $T_0$ but to find the optimal 
situation one should tune the parameters and study the border 
of the FL regime by numerical methods.  

\subsection
{Pressure dependence of the low-temperature resistivity}
The FL laws derived for the periodic Anderson model explain 
the changes observed in the transport coefficients of 
heavy fermions under applied pressure. 
As an illustration, we consider the pressure dependence of the 
coefficient of the $T^2$ term in the resistivity, defined as 
$A(p)=(\rho-\rho_0)/T^{2}$,  where $\rho_0$ is the residual resistivity.  
In the case of the two heavy fermion antiferromagnets 
CePd$_{2-x}$Ge$_{2-x}$\cite{wilhelm.02} and CeRu$_{2}$Ge$_{2}$\cite{Wilhelm.05}, 
$A(p)$ is small and nearly pressure independent for $p\leq $ 4 GPa. 
Above 4 GPa,  $A(p)$  increases rapidly and reaches a maximum 
value for pressure between 4 and 8 and GPa.  
At the critical pressure, $p_c$, the 
maximum of $A(p_c)$ is typically one order of magnitude higher 
than at ambient pressure. For $p > p_c$,  the value of $A(p)$ drops 
to the ambient pressure one.
A somewhat different behavior is found in the heavy fermion 
superconductor CeCu$_2$Si$_2$ and in the antiferromagnet CeCu$_2$Ge$_2$, 
when analyzed in the normal state\cite{jaccard.98,holmes.04}.  
The values of $A(p)$ are large at initial pressure which 
is applied in order to restore the FL behavior; 
above that pressure $A(p)$ decreases to a plateau 
and then drops by nearly two orders of 
magnitude\cite{jaccard.98,holmes.04}.

A sharp maximum of $A(p)$  is observed for systems in which 
the ground state of the 4$f$ ion is characterized at ambient pressure 
by a CF doublet well separated from the excited CF states.  
The Neel temperature of such systems is much higher than the 
Kondo temperature and large paramagnetic entropy of Ce ions 
is removed at $T_N$ by an AFM transition rather than by a Kondo 
crossover. The low-entropy state involves large unscreened 
moments which correspond to the 4$f$ ions frozen in the 
high-temperature (magnetic) configuration. 
Below $T_N$, the conduction electrons are essentially free, 
except for some magnon and impurity scattering, such that the 
resistivity at ambient pressure is small and weakly temperature-dependent, 
as shown by CePd$_{2-x}$Ge$_{2-x}$ and CeRu$_{2}$Ge$_{2}$ data. 

To explain the experimental data we assume that an 
increase of pressure leads at $p_c$ to the delocalization 
of the 4$f$ states\cite{kernavanois.2005} and the formation of 
hybridized bands, which can be described by the $SU(2)$ Anderson model. 
The two channels (sub-bands) in which the lowest 4$f$ doublet  
hybridizes with the conduction states of the appropriate symmetry, 
accommodate nearly one electron per site, $n_c+n_f/2\simeq 1$, 
for each flavor.    
The renormalized Fermi volume is large and the FS is close 
to the edge of the Brillouin zone, where $ {v^2_F} $ is small. 
The corresponding value of $\mu_L$ is of the order 
of the bandwidth [see Eqs.~\eqref{FS_0} and ~\eqref{n_0}], 
such that ${\cal N}_c^0(\mu_L)$ is also small. 
[The maximum of ${\cal N}_c^0(\omega)$ is assumed to be close 
to the center of the band.]
Since $T_0$ is small ($T_0 < T_N$), the coefficient 
of the $T^2$ term in the resistivity, given by Eq. \eqref{eq: rho_tt} 
for $N=2$, is large. 
The enhancement of $A(p)$ with respect to the values at 
ambient pressure  (where the $f$ states are localized) is due 
to the delocalization of the $f$ states by applied pressure. 
In a system with delocalized $f$ states, the main effect of 
pressure is to increase the hybridization and $T_0$, provided 
the degeneracy of the lowest occupied CF level is preserved. 
We can understand the increase of $T_0$ by recalling that 
the FL scale of the lattice is proportional to the low-energy 
scale of the auxiliary impurity model [see Eq.~\eqref{eq: tilde_omega_T0}] 
and that an increase of the Kondo scale with hybrdization (pressure) 
is a typical feature of any Kondo system. 
An increases of $T_0$ can also be inferred from the experimental 
data\cite{jaccard.98,holmes.04}, which show that in 
CePd$_{2-x}$Ge$_{2-x}$ and CeRu$_{2}$Ge$_{2}$ the value of $T_0$ 
scales with the high-temperature Kondo scale. 

A qualitative change occurs at the point where the hybridization 
becomes so large that the system cannot sustain the CF excitations  
and the degeneracy of the $f$ level changes from two to six. 
To estimate the resistivity of this high-pressure state 
we use the $SU(6)$ model in which a single $f$ electron 
is distributed over six equivalent hybridized channels. 
In that case, our FL solution shows that the FS is shifted away from 
the zone boundary,  $\mu_L$ is decreased and ${\cal N}_c^0(\mu_L)$ 
increased, with respect to the values obtained for $N=2$.  
The average squared velocity $ {v^2_F} $  and the 
FL scale $T_0$ are also increased for $N=6$, which reduces 
$A(p)$ to small values, in agreement with the experimental 
data~\cite{jaccard.98,holmes.04,wilhelm.02,Wilhelm.05}.
The drop of $A(p)$ for  $ p > p_c$ signifies the doublet-sextet 
crossover and is mainly due to the pre-factor in Eq.~\eqref{eq: rho_tt}, 
which has an explicit and implicit dependence on $N$. 
At the crossover, the Fermi volume changes and the KW ratio 
is strongly pressure dependent. 
Once the degeneracy of the $f$ state is stabilized at 
a higher value, a further increase of pressure reduces $A(p)$ 
by increasing $T_0$ but does not change  ${v^2_F} $ 
or $\mu_L$, which are fixed by the Luttinger 
theorem for $(n_c+n_f/6)$ states per channel. 
Eventually, at very high pressure, the system is transformed 
into a valence fluctuator with an enormously enhanced FL scale, 
such that $A(p)$ drops to a very small value. 

As regards CeCu$_2$Si$_2$\cite{holmes.04} and 
CeCu$_2$Ge$_2$\cite{jaccard.98}, we can explain the data 
using the same reasoning as above, if we assume that 
the {\it f} states are delocalized at the initial pressure 
which restores the FL state.
The value of $A(p)$ at the initial pressure is large, 
because the temperature dependence of the resistivity is 
due to two degenerate sub-bands with very heavy fermions.
The FS of these sub-bands is close to the zone boundary, 
where ${v^2_F}$, $\mu_L$, and ${\cal N}_c^0(\mu_L)$ are 
very small. The hybridized bands which involve the 
excited CF states have a small Fermi volume and their 
low-energy properties are free-electron like. They carry 
most of the current but have temperature-independent 
resistivity  and do not affect $A(p)$. 
The first decrease of $A(p)$ occurs at pressure at which 
an increase of the hybridization removes the CF excitations, 
such that the effective degeneracy of the $f$ state increases 
from 2 to 4 or 6. For large $N$, the expression in Eq.~\eqref{eq: rho_tt} 
gives small $A$. The final drop of $A(p)$ is due to the 
crossover into the valence fluctuating regime and an  
exponential increase of the FL scale $T_0$. 
If we assume a proportionality between $T_0$ and the Kondo 
scale $T_K$, which is indicated by the experimental data, 
our calculations would explain the $A(p)$ versus $T_K$ 
scaling reported in Ref.~\onlinecite{holmes.04}. 

The above scenario also offers an explanation for the peak observed 
in the residual resistivity of CeCu$_2$Ge$_2$~\cite{jaccard.98} 
and CeCu$_2$Si$_2$~\cite{holmes.04} for pressures at which 
the degeneracy of the $f$ level changes and $A(p)$ drops from 
the maximum value. At such a pressure, the formerly non-resonant 
channels responsible for the low residual resistivity transform 
into resonant ones, which reduces ${v^2_F} $ 
and ${\cal N}_c^0(\mu_L)$. For a constant impurity scattering 
rate, this leads to a drastic increase of $\rho_0$. 
A further increase of pressure transforms the system into 
a valence fluctuator with nearly free conduction states 
and small $\rho_0$. 
An alternative explanation for this peak has been suggested by Miyake 
and Maebashi~\cite{miyake.02}, who attribute it to critical valence 
fluctuations. In our approach, these would occur at higher
pressure, and we expect the resulting feature to be narrower 
than the one observed experimentally. 
A peak in $\rho_0(p)$ is also observed in YbCu$_2$Si$_2$~\cite{jaccard.98}. 
In this case, the mechanism that leads to it is the 
"mirror-image" of the preceding one\cite{zlatic.05}: 
starting from a valence fluctuating regime at ambient pressure, 
the system is driven into a state with a well defined valence 
in which all eight components of the $J=7/2$ $4f$-multiplet 
hybridize with conduction states, so that $\rho_0$ is very large. 
A further increase of pressure stabilizes the magnetic 4$f^{13}$ 
configuration\cite{alami-yadri-99}, inhibits the charge fluctuations,  
and gives rise to the CF excitations. This reduces the effective 
degeneracy of the $f$ hole by splitting-off the doubly-degenerate 
'resonant' sub-band with heavy QPs from the "non-resonant" sub-bands 
with nearly free electrons. For the reasons given above, $\rho_0$ is 
reduced yet larger than at ambient pressure. 

\section{Summary and conclusions 
\label{Sec: summary}} 

A detailed understanding of the thermoelectric properties of 
the periodic Anderson model in the full temperature range and 
for arbitrary parameters is of a considerable interest, 
as it might facilitate the search for new thermoelectrics 
with a useful low-temperature figure-of-merit. 
The model with the $SU(N)$ symmetry and an infinitely 
large Coulomb repulsion between $f$ electrons captures the 
low-temperature  features of the intermetallic compounds with 
Ce, Eu and Yb ions which exhibit often a large thermopower 
and might have a potential for applications. 

The thermoelectric response of the $SU(N)$ model with $n=N n_f+n_c$ 
particles is calculated in the FL regime assuming that 
the FL scale $T_0$ is known. 
The value of  $T_0$ can be obtained from the numerical DMFT 
results for the linear coefficient of the specific heat or 
estimated from an analytic expression for the QP DOS given  
by the slave-boson approximation\cite{burdin.08}. 
The transport integrals are related to the current-current 
correlation function and calculated by using the Luttinger 
theorem and the DMFT condition. The low-temperature transport 
coefficients are obtained from the Sommerfeld expansion as 
power series in terms of the reduced temperature $T/T_0$. 
The coefficients in the expansion depend on the average 
conduction electron velocity, the unit cell volume, 
the effective degeneracy of the $f$ state, 
the unrenormalized density of {\it c} states, ${\cal N}_c^0(\mu_L)$,  
and the shift in the chemical potential $\Delta\mu=\mu_L-\mu$ 
which also has a simple physical interpretation.  
The chemical potential $\mu$ corresponds to $n$ 
interacting particles and for a flat low-energy dispersion, $\mu$ is 
not much different from the chemical potential of $N n_c$ conduction 
states decoupled from the $f$ states. The FS of such a conduction band 
encloses $Nn_c$ points in the ${\bf k}$-space and is considered 
to be 'small'.
The FS of the Anderson model, determined by the Luttinger theorem,  
accommodates $f$ electrons in addition to $c$ electrons 
and is considered to be 'large'. 
For ${\bf k}$-independent hybridization, this 'large' FS coincides 
with the FS of a free conduction band with $n$ electrons and 
chemical potential $\mu_L$.  
Obviously, the FS of a non-interacting conduction band with $n$  
electrons differs from the FS of $N n_c$ electrons, 
so that the shift $\Delta\mu$ measures the 'jump' in the Fermi 
volume due to the hybridization. 

The parameter dependence of the coefficients multiplying 
various powers of reduced temperature corrects the simple scaling 
behavior and explains the deviations of the KW and the $q$-ratio 
from the universal constants. 
The FL law for the conductivity gives the KW ratio which depends 
on the multiplicity of the $f$ state,  the unit cell volume,
the average FS velocity, and the carrier concentration, 
in agreement with the experimental data~\cite{hussey.2005}. 
The FL law for the thermopower gives the $q$-ratio which 
depends on the concentration of $f$ electrons. 
From these results we conclude that, in some systems, 
pressure or chemical pressure can cause a substantial 
shift of the 'universal ratios' from the common values. 
We also find that the quasiparticle damping leads to the break-down 
of the WF law due to the temperature dependence of the effective 
Lorenz number. In the absence of the thermal current due to phonons, 
this would lead to a substantial enhancement of the thermoelectric 
figure-of-merit $ ZT>1$.
Assuming that pressure changes the hybridization and the 
effective degeneracy of the $f$ state we explained the pronounced 
maximum of the coefficient of the $T^2$ term of the electrical 
resistance~\cite{jaccard.98,holmes.04,wilhelm.02,Wilhelm.05} 
and the rapid variation of residual resistance~\cite{holmes.04,jaccard.98} 
found in a number of Ce and Yb intermetallics at some critical pressure.  

Our results describe the main features of thermal transport 
that one finds in heavy fermions in the FL regime but do not 
indicate the temperature at which the FL description breaks 
down nor relate the FL scale to the Kondo scale, 
which characterizes the high-temperature behavior. 
A complete discussion should explain the low- and high-temperature 
behavior on equal footing, provide all the relevant energy 
scales of the system, and account for the change in the 
effective degeneracy of the low-energy states due to 
applied pressure, doping or temperature.  
The change of the effective degeneracy can have a dramatic effect 
on the transport coefficients and seems to be responsible for 
complicated thermoelectric response of intermetallic compounds 
with Ce and Yb ions. 
The realistic modeling should remove the $SU(N)$ symmetry, 
take into account the excited CF states, consider the details 
of the band structure, and/or include aditional interactions. 
Unfortunately, none of the presently available methods can 
solve such more realistic models, explain the crossovers between 
various physical states, and describe the behavior of the correlation 
functions in the full temperature and pressure range.

A rough description of the high-temperature regime, which takes 
into account  the CF splitting, can be obtained by assuming that the 
conduction electrons scatter incoherently on the 4{\it f} ions. 
This reduces the lattice model with $n$ electrons per unit cell 
to an effective Anderson impurity model which can be solved 
very accurately by the NCA~\cite{zlatic.05,zlatic.07b}. 
The NCA solution leads to an exponentially small Kondo scale 
and explains the main features that one finds in the thermoelectric 
response of heavy fermions and valence fluctuators for $T\geq T_K$. 
In particular, the NCA shows that the effective degeneracy of the model 
changes as the excited CF become thermally populated. 
Some justification for applying an effective single impurity model 
to the stoichiometric compounds is provided by the fact 
that in the temperature range we are concerned with 
the resistivity of most heavy fermions is very large and  
the mean free path is not much longer than a few lattice spacings.  
Furthermore, the NCA solution for an effective two-fold degenerate impurity 
model~\cite{zlatic.05,zlatic.07b}  agrees for $T\geq T_K$ 
with recent DMFT+NRG results for the two-fold degenerate periodic 
Anderson model~\cite{grenzebach.06}. 
For an $N$-fold degenerate model the NCA gives $\alpha(T)$ and 
the power factor $P(T)=\alpha^2(T)/\rho(T)$ with a pronounced 
maximum and ${\cal L}(T)$ with a shallow minimum around $T_K$, 
which supports our previous conclusions regarding the enhancement 
of the thermoelectric figure-of-merit. 
The agreement between the perturbative solution of an effective 
single impurity model and the experimental data 
indicates that the high-temperature state of heavy fermions 
and valence fluctuators can be represented by a nearly free 
conduction band which is weakly perturbed by localized 
(paramagnetic) $f$ states. 
Ignoring the thermal broadening of the Fermi distribution function 
we find that this FS is much smaller than the low-temperature 
one which must include the $f$ electrons in order to satisfy 
the Luttinger theorem. 
The perturbative solution breaks down in the coherent regime, 
as it cannot describe the change in the Fermi volume.
For a periodic model, the reduction of the paramagnetic entropy, 
the crossover from the high-temperature perturbative regime into 
the FL regime, and relationship between $T_K$ and $T_0$ can only 
be obtained by non-perturbative methods. 

The DMFT+NRG method describes, in principle,  the crossover 
from the high-entropy state formed above $T_K$ to the low-entropy 
FL ground state state. It provides the Kondo scale $T_K$ which 
governs the high-temperature behavior and gives an accurate 
numerical estimate of the FL scale $T_0$ and other thermodynamic 
quantities, like the number of particles or the chemical potential 
of the ground state. 
However, neither the DMFT+NRG nor the effective impurity approach 
provide a quantitative description of the transport properties 
of the periodic Anderson model for temperatures below $T_K$ or $T_0$. 
The usefulness of the FL approach is that it gives the 
low-temperature transport coefficients for a given $T_0$ 
and allows us to obtain an overall description by 
extrapolating between the FL solution and the high-temperature 
one obtained by the NCA or the DMFT+NRG methods. 

It would be interesting to perform an experimental study 
of the pressure-induced deviations of the $q$-ratio and the KW 
ratio from their universal values, using the universal behavior of 
the power factor or the effective Lorenz number as consistency checks.
Since $P(T)$ and ${{{\cal L}}}(T)$ require only transport
measurements, they are well suited for pressure experiments.
The above discussion makes clear that pressure
experiments provide the most stringent test of the FL  laws.

\acknowledgments

We thank David Logan for pointing out some unclear points 
in the original manuscript. Useful discussion with K. Yamada, 
D. Jaccard and T. Pruschke are gratefully acknowledged.
This work has been supported by the Ministry of Science of Croatia
(Grant No. 035-0352843-2849), the COST P-16 ECOM project,
 and the National Science Foundation under Grant Nos.~DMR-0210717
 and DMR-0705266. The work at SISSA has been supported by the 
Central European Initiative. V.Z. and J.K.F acknowledge the 
 hospitality of the ETH Zurich, where this work has been completed. 

\appendix

\section{Correlation function in the Fermi liquid regime
\label{correlation_function}}
To find $\Lambda(\omega,T)$  we use the Kubo formula for the 
static conductivity which provides $L_{11}$ as the zero-frequency 
limit of the number current -number current correlation 
function\cite{mahan.81} 
\begin{eqnarray}
L_{11}^{\alpha\beta}
&=&\lim_{\nu\rightarrow 0}{\rm Re}\frac{i}{\nu}\bar L_{11}^{\alpha\beta}(\nu),
\label{eq: l11_0}
\end{eqnarray}
where 
\begin{eqnarray}
\bar L_{11}^{\alpha\beta}(i\nu_l)
&=&
\frac{\hbar}{{\cal V}{\cal N}_i}}{\int_0^{\beta}d\tau e^{i\nu_l\tau}
\langle T_\tau {\bf j}_{\alpha}^\dagger(\tau){\bf j}_{\beta}(0)\rangle,
\label{eq: l11}
\end{eqnarray}
$\nu_l=2\pi k_B T l $ is the Bosonic Matsubara frequency, 
the $\tau$-dependence of the operator is with respect to the 
Hamiltonian in Eq.~\eqref{Hamiltonian},
the subscripts $\alpha$ and $\beta$ denote the respective spatial 
indices of the number-current vector ${\bf j}_c/e$, 
and we must analytically continue 
$\bar L_{11}^{\alpha\beta}(i\nu_l)$ to the real
axis $\bar L_{11}^{\alpha\beta}(\nu)$ before taking the limit 
$\nu\rightarrow 0$.  

Substituting the definition of the particle current operator 
from Eq.~\eqref{current} into Eq.~\eqref{eq: l11} for $\bar L_{11}^{\alpha\beta}$ 
and taking the limit of infinite dimensions, in which the dressed 
correlation function is equal to the bare one
\cite{khurana.1990,zlatic-horvatic.1990}, 
we obtain 
\begin{equation}
\bar L_{11}^{\alpha\beta}(i\nu_l)
=
\frac{-\hbar}{{\cal V}{\cal N}_i}\int_0^\beta d\tau e^{i\nu_l\tau}
\sum_{\bf k}{\bf v}_{{\bf k}\alpha}{\bf v}_{{\bf k}\beta}
G_{c}({\bf k},\tau)G_{c}({\bf k},-\tau).
\label{eq: l11_bare}
\end{equation}
where $G_c({\bf k},\tau)$ is the imaginary time 
Green's function of the {\it c} electrons, which 
can be expressed as a Fourier series  
$G_c({\bf k},\tau)=k_BT \sum_n\exp(-i\omega_n\tau)G_c({\bf k},i\omega_n)$.  
Substituting into Eq.~(\ref{eq: l11_bare}) and integrating over 
imaginary time provides the result 
\begin{eqnarray}
                  \nonumber
\bar L_{11}^{\alpha\beta}(i\nu_l)
=
\frac{-\hbar}{{\cal V}{\cal N}_i}k_B T\sum_n
\sum_{\bf k}{\bf v}_{{\bf k}\alpha}{\bf v}_{{\bf k}\beta}\\
G_{c}({\bf k},i\omega_n)G_{c}({\bf k},i\omega_n+i\nu_l) 
                \label{eq: l11_imag}
\end{eqnarray}
which has to be analytically  continued to the real axis. 
Since the Green's functions in Eq.~\eqref{eq: l11_imag} 
depend on ${\bf k}$ only through $\epsilon({\bf k})$ 
which is an even function of ${\bf k}$, while ${\bf v}_{{\bf k}\alpha}$ 
is odd, the summation over ${\bf k}$ vanishes for $\alpha\neq \beta$. 
We consider only isotropic systems, where
$L_{11}^{\alpha\alpha}=L_{11}$. 
To perform the analytic continuation we follow closely 
Ref.~\onlinecite{freericks.01} and obtain the result 
(evaluated explicitly for a three-dimensional system)
\begin{eqnarray}
&&L_{11}
=
\frac{-\hbar}{{\cal V}{\cal N}_i}
\frac{{v^2_F} }{3\pi}
\int_{-\infty}^{\infty}d\omega 
\lim_{\nu\rightarrow 0}
\frac{f(\omega)-f(\omega+\nu)}{\nu}  \label{eq: l11_real3}\\
&\times&
{\rm Re} \biggr [
\frac{G_c(\omega)-G_c(\omega+\nu)}
{\nu+\Sigma_c(\omega)-\Sigma_c(\omega+\nu)}
-
\frac{G_c^*(\omega)-G_c(\omega+\nu)}
{\nu+\Sigma_c^*(\omega)-\Sigma_c(\omega+\nu)}
\biggr ].
\nonumber
\end{eqnarray}
where we introduced the local Green's function 
\begin{equation}
                             \label{eq: Gc_local}
G_c(\omega)
=
\sum_{\bf k}  G_c({\bf k},\omega) ,
\end{equation}
and replaced the square of the $\alpha$-component of 
velocity by its Fermi surface average ${v^2_F} /d$
($d$ is the spatial dimension, which we can take to be 
equal to 3 for real systems). 
This step is justified, because the energy integrations 
in Eq.~\eqref{eq: l11_real3} is restricted to a narrow interval 
around the Fermi energy, where the integrand is singular, 
so that the main contribution to the ${\bf k}$ summation 
comes from the ${\bf k}$-points close to the Fermi surface;
on the infinite-dimensional hypercubic lattice, the integral
can be performed exactly and one finds that the average
square velocity is equal to $a_l^2t^{*2}/16\hbar^2$ when expressed
in terms of the reduced nearest-neighbor hopping $t^*=2\sqrt{d}t$. 

We can now take the limit of $\nu\rightarrow 0$. Writing 
\begin{equation}
\lim_{\nu\rightarrow 0}
\frac{G_c(\omega)-G_c(\omega+\nu)}
{\nu+\Sigma_c(\omega)-\Sigma_c(\omega+\nu)}
=
-\frac{\partial G_c(\omega) }{\partial\omega}
\frac{1}{1-\frac{\partial \Sigma_c(\omega) }{\partial\omega}} 
\label{eq: g_deriv}
\end{equation}
and 
\begin{equation}
\lim_{\nu\rightarrow 0}
\frac{G^*_c(\omega)-G_c(\omega+\nu)}
{\nu+\Sigma^*_c(\omega)-\Sigma_c(\omega+\nu)}
=
 \frac{{\mathrm Im} G_c(\omega)}{{\mathrm Im} \ \Sigma_c(\omega)}
\label{eq: im_G_Im_sigma}
\end{equation}
produces our final result 
\begin{equation}
L_{11}=\int_{-\infty}^{\infty}d\omega
\left ( -\frac{df(\omega)}{d\omega} \right )  \Lambda(\omega,T),
\label{eq: l11_final}
\end{equation}
where $\Lambda(\omega,T)$ is defined by
\begin{equation}
\Lambda(\omega)
=
\frac{\hbar} { {\cal V} {\cal N}_c}
\frac{{v^2_F} }{3\pi}
\left(\frac{{\rm Im}G_c(\omega)}{{\rm Im}\Sigma_c(\omega)}
+{\rm Re} 
\left[  
  \frac{
   \frac{\partial G_c(\omega)}{\partial\omega} }
        { 1-\frac{\partial \Sigma_c(\omega)}{\partial\omega} } 
\right]\right).
                \label{eq: lambda_FL.1}
\end{equation}
To estimate the relative importance of the two terms in  
Eq.~\eqref{eq: lambda_FL.1} 
we introduce the Hilbert transform of ${\cal N}_c(\omega)$, 
\begin{equation}
{\cal H}_c(\omega) 
= - \frac{1}{\pi{{\cal V}{\cal N}_i}}{\mathrm Re} G_c(\omega) .
\end{equation}
In the FL regime, where ${\cal N}_c(\omega) $ is $\delta$-function like, 
the slope of ${\cal H}_c(\omega)$ is very large; it is proportional 
to the $c$ electron enhancement factor
$$
{\mathrm Re} \left[
1-\frac{\partial \Sigma_c(\omega) }{\partial\omega} \right]
=
Z_c^{-1}. 
$$
On the other hand, 
${\mathrm Im} \ {\partial \Sigma_c(\omega) }/{\partial\omega} $
is small around $\omega=0$, because ${\mathrm Im} \Sigma_c(\omega) $  
is close to its maximum value. Using
$$
{\mathrm Im}\ \frac{\partial \Sigma_c(\omega) }{\partial\omega} \ll Z_c^{-1}. 
$$
we neglect 
$[{\mathrm Im} \ {\partial \Sigma_c(\omega) }/{\partial\omega} ]^2$
in the denominator of the second term for $\Lambda(\omega)$ 
in Eq.~\eqref{eq: lambda_FL.1} and approximate 
$$
{\rm Re} 
\left[  
  \frac{
   \frac{\partial G_c(\omega)}{\partial\omega} }
        { 1-\frac{\partial \Sigma_c(\omega)}{\partial\omega} } 
\right]
\simeq 
Z_c \frac{ \partial {\cal H}_c(\omega)}{\partial \omega}
-
\frac{ \partial {\cal N}_c(\omega)}{\partial \omega}
{\mathrm Im}\ \frac{\partial \Sigma_c(\omega) }{\partial \omega}.
$$
This term is small with respect to 
${ {\cal N}_c(\omega)}/{{\rm Im}\Sigma(\omega)}$ which diverges 
in the limit $T,\omega\to 0$. Keeping only the singular term 
in Eq.~\eqref{eq: lambda_FL.1} we obtain the result of
Eq.~\eqref{lambda-FL} used in the text.  

\section{Transport coefficients
\label{Transport_coefficients}}

The transport coefficients are obtained by expanding the derivative of 
the Fermi function in a Sommerfeld expansion,
$-\partial f(\omega)/\partial \omega
=\delta(\omega) + 
(\pi^2 k_B^2 T^2/6) [\partial^2 \delta(\omega)/\partial \omega^2]$, 
which gives the following result for the transport integrals 
\begin{eqnarray}
                             \label{L_mn}
L_{mn}
&=&
\left[
{\omega}^{m+n-2} \Lambda(\omega,T)
\right]_{\omega=0} \\
&+&
\frac{\pi^2 k_B^2 T^2 }{6}
\left\{
\frac{ \partial^2} {\partial\omega^2}
\left[ \omega^{m+n-2} \Lambda(\omega,T) \right]
\right\}_{\omega=0} .  
\nonumber
\end{eqnarray}
For thermal transport we have to evaluate 
\begin{eqnarray}
                             \label{L.11}
L_{11}
&=&
\left[
\Lambda(\omega,T)
\right]_{\omega=0}
+
\frac{\pi^2 k_B^2 T^2}{6}
\Lambda^{\prime\prime}(\omega,T)|_{\omega=0}, 
\\
                             \label{L.12}
L_{12}
&=&
\frac{\pi^2 k_B^2 T^2 }{3}
\Lambda^{\prime}(\omega,T)|_{\omega=0}, 
\\
                            \label{L_22}
L_{22}
&=&
\frac{\pi^2 k_B^2 T^2}{3}
\Lambda(\omega,T)|_{\omega=0}, 
\end {eqnarray}
where 
\begin{eqnarray}
                             \label{lambda}
\Lambda(\omega,T)
&=&
\frac{1}{3}{v^2_F}  {\cal N}_c(\omega) \tau(\omega,T) 
\\[1ex]
\Lambda^\prime(\omega,T)  \label{lambdaprime}
&=&
\frac{1}{3} {v^2_F}  
[{\cal N}^\prime_c(\omega) \tau(\omega,T) 
+
{\cal N}_c(\omega) \tau^\prime(\omega,T)] \nonumber\\
\\[1ex]
\Lambda^{\prime\prime}(\omega,T)
&=&
\frac{1}{3} {v^2_F}            
[{\cal N}^{\prime\prime}_c(\omega) \tau(\omega,T)\nonumber \\  
&+& 
2{\cal N}^\prime_c(\omega) \tau^\prime(\omega,T)
+
{\cal N}_c(\omega) \tau^{\prime \prime}(\omega,T)] , 
\label{lambda2prime}
\end {eqnarray}
and $\Lambda^\prime$ and $\Lambda^{\prime\prime}$ 
denote the first and second derivatives with respect to $\omega$. 
The first derivative of $\mathcal{N}_c(\omega)$ is 
\begin{equation}
                   \label{eq: first derivative}
\mathcal{N}_c^\prime(\omega)
=
\frac{2 (\omega-\tilde\omega_f)}{Z_f\tilde V^2}
\mathcal{N}_f(\omega)
+
\frac{(\omega-\tilde \omega_f)^2}{Z_f\tilde V^2}
\mathcal{N}_f^\prime(\omega), 
\end{equation}
and becomes in the $\omega\to 0$ limit 
\begin{eqnarray}
                                \label{eq: Nc_prime_a}
\mathcal{N}_c^\prime(0)
&\simeq&
-2\frac{\mathcal{N}_c(0)}{\tilde\omega_f},  
\end {eqnarray}
where the DMFT condition showed that the derivative 
of the $f$-electron DOS is small at $\omega=0$.
For the second derivative, we obtain 
\begin{equation}
                   \label{eq: second derivative 2}
\mathcal{N}_c^{\prime\prime}(0)
=
\frac{2 \mathcal{N}_c(0)}{\tilde\omega_f^2}, 
\end{equation} 
using 
${\mathcal{N}_f}(0) / {Z_f\tilde V^2}
=
{\mathcal{N}_c}(0) / {\tilde \omega_f^2}$ 
and dropping the terms proportional to 
 $\tilde\omega_f\mathcal{N}_f^\prime(0)$ 
and $\tilde\omega_f^2\mathcal{N}_f^{\prime\prime}(0)$ 
which are exponentially small 
[see also Eq.~\eqref{eq: Nf_prime}].

The transport relaxation time given by the expression in 
Eq. \eqref{eq: tau_low_T_omega} yields at $\omega=0$ 
\begin{eqnarray}
\tau^\prime(0,T) 
                        \label{eq: tau_prime}
=
-2\frac{\tau_0(T)}{\tilde\omega_f} ,
 \\
\tau^{\prime\prime}(0,T) 
=
2\tau_0(T)\left[\frac{1}{\tilde\omega_f^2}-\frac{1}{(\pi k_B T)^2}\right].
                        \label{eq: tau2prime}
\end{eqnarray}
The transport integrals are now easy to find. 
The first one is obtained from Eq. \eqref{lambda2prime}, 
which gives 

\begin{eqnarray}
&&L_{11}
\simeq
\frac{1}{3}{{v^2_F} \cal N}_c(0)\tau_0(T) 
\biggr\{
\left\{ 1 + \right. 
\biggr .
\nonumber\\
&+&
          \nonumber 
\biggr.
\frac{\pi^2 k_B^2 T^2}{3}
\left[
     \frac{6\mathcal{N}_c(0)}{\tilde\omega_f^2}
    -\frac{1}{(\pi k_BT)^2}
\right]
\biggr\} .
\nonumber\\[2ex]
&\simeq&
                             \label{lambda_tt^2}
{\Lambda(0,T)}\left[ \frac{2}{3}+{\cal O}(T^2)\right].
\end {eqnarray}
The last term in the square bracket in the second line grows as $1/T^2$, 
while the first one is a (large) constant which we  neglect 
at low enough temperature. 
Since ${\Lambda(0,T)}\propto 1/T^2$, this approximation 
amounts to keeping the $T^2$ terms in the electrical 
resistance and neglecting the $T^4$ contribution. 
The correcting factor 2/3 in  Eq. \eqref{lambda_tt^2} originates 
from the imaginary part of the self energy and it is 
well known from the dilute alloy problem\cite{horvatic.84}; hence it arises only in the clean limit. 
Inserting the expressions for the renormalized c-DOS [Eq.\eqref{eq:N_c_omega}]
and the relaxation time [Eq.\eqref{eq: tau_00}], we obtain the
dominant low-temperature contribution to $L_{11}$: 
\begin{equation}
                             \label{lambda_tt}
L_{11}
=
{{{v^2_F} }}
\frac{\hbar(N-1) [N {\cal V} \mathcal{N}_c^0(\mu_L)]^2}
{9\pi^3 {\cal V} }\left(\frac{ T_0}{T}\right)^2 
\end{equation}
which yields for the static conductivity $\sigma(T)=Ne^2 L_{11}$
\begin{equation}
                             \label{sigma_tt}
\sigma(T)
=
\frac{\hbar{e^2{{v^2_F} }}
N(N-1)[N{\cal V}\mathcal{N}_c^0(\mu_L)]^2}
{9\pi^3 {\cal V}} 
\left(\frac{ T_0}{T}\right)^2. 
\end{equation}
and the resistivity $\rho(T)=1/\sigma(T)$. 

The second transport integral is obtained from 
Eq.~\eqref{lambdaprime} which gives 
\begin{eqnarray}
L_{12}
&=&
\frac{\pi^2 k_B^2 T^2 }{3} 
\Lambda(0,T) \\
&\times&
\left[\frac{{\cal N}^\prime_c(0)}{{\cal N}_c(0)} 
+
\frac{\tau^\prime(0,T)}{\tau(0,T)}\right]\nonumber\\
&\simeq& 
- 2L_{11}
\frac{\pi^2 k_B^2 T^2 }{\tilde\omega_f} 
\simeq
                                  \label{eq: L.12_3}
\mp 4 L_{11} \frac{\pi^2 k_B T^2}{n_f T_0} . 
\end {eqnarray}
 In the second equation we used 
$\tilde\omega_f\simeq \pm n_f k_B T_0/2 $,  
which holds for heavy fermions with $n_f\simeq 1$ 
and large $N$ only.

The thermal conductivity is defined by the expression, 
\begin{equation}
                                                  \label{kappa1}
\kappa(T)
=
N\frac{L_{11}}{T}
\left [\frac{L_{22}}{L_{11}}-\left(\frac{L_{12}}{L_{11}}\right)^2
\right] 
\end {equation}
and follows from the previous results which give 
$L_{11}=2\Lambda(0,T)/3$, 
${L_{12}}/L_{11}\simeq {2\pi^2 k_B^2 T^2 }/{\tilde\omega_f} $,  
$L_{22}={\pi^2 k_B^2 T^2} \Lambda(0,T)/3$, and 
${L_{22}}/{L_{11}}={\pi^2 k_B^2 T^2}/{2}$.
In the heavy fermion limit this yields the expression in  
Eq.~\eqref{kappa2} used in the main text. 

\bibliographystyle{apsrev}
\bibliography{zlatic_PRB_resubmitted}

\begin{thebibliography}{36}
\expandafter\ifx\csname natexlab\endcsname\relax\def\natexlab#1{#1}\fi
\expandafter\ifx\csname bibnamefont\endcsname\relax
  \def\bibnamefont#1{#1}\fi
\expandafter\ifx\csname bibfnamefont\endcsname\relax
  \def\bibfnamefont#1{#1}\fi
\expandafter\ifx\csname citenamefont\endcsname\relax
  \def\citenamefont#1{#1}\fi
\expandafter\ifx\csname url\endcsname\relax
  \def\url#1{\texttt{#1}}\fi
\expandafter\ifx\csname urlprefix\endcsname\relax\def\urlprefix{URL }\fi
\providecommand{\bibinfo}[2]{#2}
\providecommand{\eprint}[2][]{\url{#2}}

\bibitem[{\citenamefont{Behnia et~al.}(2004)\citenamefont{Behnia, Jaccard, and
  Flouquet}}]{behnia.04}
\bibinfo{author}{\bibfnamefont{K.}~\bibnamefont{Behnia}},
  \bibinfo{author}{\bibfnamefont{D.}~\bibnamefont{Jaccard}}, \bibnamefont{and}
  \bibinfo{author}{\bibfnamefont{J.}~\bibnamefont{Flouquet}},
  \bibinfo{journal}{J. Phys.: Condens. Matter} \textbf{\bibinfo{volume}{16}},
  \bibinfo{pages}{5187} (\bibinfo{year}{2004}).

\bibitem[{\citenamefont{Sakurai and Isikawa}(2005)}]{sakurai.05}
\bibinfo{author}{\bibfnamefont{J.}~\bibnamefont{Sakurai}} \bibnamefont{and}
  \bibinfo{author}{\bibfnamefont{Y.}~\bibnamefont{Isikawa}},
  \bibinfo{journal}{J. Phys. Soc. Japan} \textbf{\bibinfo{volume}{74}},
  \bibinfo{pages}{1926} (\bibinfo{year}{2005}).

\bibitem[{\citenamefont{Hossain et~al.}(2004)\citenamefont{Hossain, Geibel,
  Senthilkumaran, Deppe, Baenitz, Schiller, and Molodtsov}}]{hossain.04}
\bibinfo{author}{\bibfnamefont{Z.}~\bibnamefont{Hossain}},
  \bibinfo{author}{\bibfnamefont{C.}~\bibnamefont{Geibel}},
  \bibinfo{author}{\bibfnamefont{N.}~\bibnamefont{Senthilkumaran}},
  \bibinfo{author}{\bibfnamefont{M.}~\bibnamefont{Deppe}},
  \bibinfo{author}{\bibfnamefont{M.}~\bibnamefont{Baenitz}},
  \bibinfo{author}{\bibfnamefont{F.}~\bibnamefont{Schiller}}, \bibnamefont{and}
  \bibinfo{author}{\bibfnamefont{S.~L.} \bibnamefont{Molodtsov}},
  \bibinfo{journal}{Phys. Rev. B} \textbf{\bibinfo{volume}{69}},
  \bibinfo{pages}{014422} (\bibinfo{year}{2004}).

\bibitem[{\citenamefont{Sakurai et~al.}(2002)\citenamefont{Sakurai, Iwasaki,
  Lu, Ho, Isikawa, Fern\'andez, and Sal}}]{sakurai.02}
\bibinfo{author}{\bibfnamefont{J.}~\bibnamefont{Sakurai}},
  \bibinfo{author}{\bibfnamefont{A.}~\bibnamefont{Iwasaki}},
  \bibinfo{author}{\bibfnamefont{Q.}~\bibnamefont{Lu}},
  \bibinfo{author}{\bibfnamefont{D.}~\bibnamefont{Ho}},
  \bibinfo{author}{\bibfnamefont{Y.}~\bibnamefont{Isikawa}},
  \bibinfo{author}{\bibfnamefont{J.~R.} \bibnamefont{Fern\'andez}},
  \bibnamefont{and} \bibinfo{author}{\bibfnamefont{C.~G.} \bibnamefont{Sal}},
  \bibinfo{journal}{J. Phys. Soc. Japan} \textbf{\bibinfo{volume}{71}},
  \bibinfo{pages}{2829} (\bibinfo{year}{2002}).

\bibitem[{\citenamefont{O\v{c}ko et~al.}(2001)\citenamefont{O\v{c}ko, Drobac,
  Sarrao, and Fisk}}]{ocko.01}
\bibinfo{author}{\bibfnamefont{M.}~\bibnamefont{O\v{c}ko}},
  \bibinfo{author}{\bibfnamefont{D.}~\bibnamefont{Drobac}},
  \bibinfo{author}{\bibfnamefont{J.~L.} \bibnamefont{Sarrao}},
  \bibnamefont{and} \bibinfo{author}{\bibfnamefont{Z.}~\bibnamefont{Fisk}},
  \bibinfo{journal}{Phys. Rev.} \textbf{\bibinfo{volume}{64}},
  \bibinfo{pages}{085103} (\bibinfo{year}{2001}).

\bibitem[{\citenamefont{{O\v cko} et~al.}(2004)\citenamefont{{O\v cko}, Sarrao,
  and \v~Z.~\v Simek}}]{ocko.04}
\bibinfo{author}{\bibfnamefont{M.}~\bibnamefont{{O\v cko}}},
  \bibinfo{author}{\bibfnamefont{J.~L.} \bibnamefont{Sarrao}},
  \bibnamefont{and} \bibinfo{author}{\bibnamefont{\v~Z.~\v Simek}},
  \bibinfo{journal}{J. Mag. Mag. Mater.} \textbf{\bibinfo{volume}{43 - 46}},
  \bibinfo{pages}{284} (\bibinfo{year}{2004}).

\bibitem[{\citenamefont{Kohler et~al.}(2008)\citenamefont{Kohler, Oeschler,
  Steglich, Maquilon, and Fisk}}]{kohler.2008}
\bibinfo{author}{\bibfnamefont{U.}~\bibnamefont{Kohler}},
  \bibinfo{author}{\bibfnamefont{N.}~\bibnamefont{Oeschler}},
  \bibinfo{author}{\bibfnamefont{F.}~\bibnamefont{Steglich}},
  \bibinfo{author}{\bibfnamefont{S.}~\bibnamefont{Maquilon}}, \bibnamefont{and}
  \bibinfo{author}{\bibfnamefont{Z.}~\bibnamefont{Fisk}},
  \bibinfo{journal}{Physical Review B} \textbf{\bibinfo{volume}{77}},
  \bibinfo{pages}{104412} (\bibinfo{year}{2008}).

\bibitem[{\citenamefont{Kadowaki and Woods}(1987)}]{kadowaki.87}
\bibinfo{author}{\bibfnamefont{K.}~\bibnamefont{Kadowaki}} \bibnamefont{and}
  \bibinfo{author}{\bibfnamefont{S.~B.} \bibnamefont{Woods}},
  \bibinfo{journal}{Solid State Commun.} \textbf{\bibinfo{volume}{71}},
  \bibinfo{pages}{1149} (\bibinfo{year}{1987}).

\bibitem[{\citenamefont{Kontani}(2004)}]{kontani.04}
\bibinfo{author}{\bibfnamefont{H.}~\bibnamefont{Kontani}}, \bibinfo{journal}{J.
  Phys. Soc. Japan} \textbf{\bibinfo{volume}{73}}, \bibinfo{pages}{515}
  (\bibinfo{year}{2004}).

\bibitem[{\citenamefont{Tsujii et~al.}(2005)\citenamefont{Tsujii, Kontani, and
  Yoshimura}}]{KW-experiment}
\bibinfo{author}{\bibfnamefont{N.}~\bibnamefont{Tsujii}},
  \bibinfo{author}{\bibfnamefont{H.}~\bibnamefont{Kontani}}, \bibnamefont{and}
  \bibinfo{author}{\bibfnamefont{K.}~\bibnamefont{Yoshimura}},
  \bibinfo{journal}{Phys. Rev. Lett.} \textbf{\bibinfo{volume}{94}},
  \bibinfo{pages}{057201} (\bibinfo{year}{2005}).

\bibitem[{\citenamefont{Hussey}(2005)}]{hussey.2005}
\bibinfo{author}{\bibfnamefont{N.~E.} \bibnamefont{Hussey}},
  \bibinfo{journal}{J. Phys. Soc. Japan} \textbf{\bibinfo{volume}{74}},
  \bibinfo{pages}{1107} (\bibinfo{year}{2005}).

\bibitem[{\citenamefont{Yamada and Yosida}(1986)}]{yamada.86}
\bibinfo{author}{\bibfnamefont{K.}~\bibnamefont{Yamada}} \bibnamefont{and}
  \bibinfo{author}{\bibfnamefont{K.}~\bibnamefont{Yosida}},
  \bibinfo{journal}{Prog. Theor. Phys.} \textbf{\bibinfo{volume}{76}},
  \bibinfo{pages}{681} (\bibinfo{year}{1986}).

\bibitem[{\citenamefont{Yamada}(2004)}]{yamada.04}
\bibinfo{author}{\bibfnamefont{K.}~\bibnamefont{Yamada}},
  \emph{\bibinfo{title}{Electron Correlation in Metals}}
  (\bibinfo{publisher}{Cambridge University Press},
  \bibinfo{address}{Cambridge}, \bibinfo{year}{2004}).

\bibitem[{\citenamefont{Miyake and Kohno}(2005)}]{miyake.05}
\bibinfo{author}{\bibfnamefont{K.}~\bibnamefont{Miyake}} \bibnamefont{and}
  \bibinfo{author}{\bibfnamefont{H.}~\bibnamefont{Kohno}}, \bibinfo{journal}{J.
  Phys. Soc. Japan} \textbf{\bibinfo{volume}{74}}, \bibinfo{pages}{254}
  (\bibinfo{year}{2005}).

\bibitem[{\citenamefont{Mott and Jones}(1958)}]{mott-jones}
\bibinfo{author}{\bibfnamefont{N.~F.} \bibnamefont{Mott}} \bibnamefont{and}
  \bibinfo{author}{\bibfnamefont{H.}~\bibnamefont{Jones}},
  \emph{\bibinfo{title}{The Theory of the Properites of Metals and Alloys}}
  (\bibinfo{publisher}{Dover Publications}, \bibinfo{address}{London},
  \bibinfo{year}{1958}).

\bibitem[{\citenamefont{Grenzebach et~al.}(2006)\citenamefont{Grenzebach,
  Anders, Czycholl, and Pruschke}}]{grenzebach.06}
\bibinfo{author}{\bibfnamefont{C.}~\bibnamefont{Grenzebach}},
  \bibinfo{author}{\bibfnamefont{F.~B.} \bibnamefont{Anders}},
  \bibinfo{author}{\bibfnamefont{G.}~\bibnamefont{Czycholl}}, \bibnamefont{and}
  \bibinfo{author}{\bibfnamefont{T.}~\bibnamefont{Pruschke}},
  \bibinfo{journal}{Phys. Rev. B} \textbf{\bibinfo{volume}{74}},
  \bibinfo{pages}{195119} (\bibinfo{year}{2006}).

\bibitem[{\citenamefont{Georges et~al.}(1996)\citenamefont{Georges, Kotliar,
  Krauth, and Rozenberg}}]{georges-kotliar-krauth-rozenberg.1996}
\bibinfo{author}{\bibfnamefont{A.}~\bibnamefont{Georges}},
  \bibinfo{author}{\bibfnamefont{G.}~\bibnamefont{Kotliar}},
  \bibinfo{author}{\bibfnamefont{W.}~\bibnamefont{Krauth}}, \bibnamefont{and}
  \bibinfo{author}{\bibfnamefont{M.~J.} \bibnamefont{Rozenberg}},
  \bibinfo{journal}{Rev. Mod. Phys.} \textbf{\bibinfo{volume}{68}},
  \bibinfo{pages}{13} (\bibinfo{year}{1996}).

\bibitem[{\citenamefont{Zlati\'c et~al.}(2007)\citenamefont{Zlati\'c, Monnier,
  Freericks, and Becker}}]{zlatic.07b}
\bibinfo{author}{\bibfnamefont{V.}~\bibnamefont{Zlati\'c}},
  \bibinfo{author}{\bibfnamefont{R.}~\bibnamefont{Monnier}},
  \bibinfo{author}{\bibfnamefont{J.}~\bibnamefont{Freericks}},
  \bibnamefont{and} \bibinfo{author}{\bibfnamefont{K.~W.}
  \bibnamefont{Becker}}, \bibinfo{journal}{Phys. Rev. B}
  \textbf{\bibinfo{volume}{76}}, \bibinfo{pages}{085122}
  (\bibinfo{year}{2007}).

\bibitem[{\citenamefont{Hewson}(1993)}]{hewson}
\bibinfo{author}{\bibfnamefont{A.}~\bibnamefont{Hewson}},
  \emph{\bibinfo{title}{The Kondo Problem to Heavy Fermions}}
  (\bibinfo{publisher}{Cambridge University Press},
  \bibinfo{address}{Cambridge}, \bibinfo{year}{1993}).

\bibitem[{\citenamefont{Ehrenreich and Spaepen}(1997)}]{mahan.98}
\bibinfo{author}{\bibfnamefont{H.}~\bibnamefont{Ehrenreich}} \bibnamefont{and}
  \bibinfo{author}{\bibfnamefont{F.}~\bibnamefont{Spaepen}},
  \emph{\bibinfo{title}{Good Thermoelectrics}} (\bibinfo{publisher}{Academic
  Press}, \bibinfo{address}{San Diego}, \bibinfo{year}{1997}),
  vol.~\bibinfo{volume}{51} of \emph{\bibinfo{series}{Solid State Physics}},
  chap. \bibinfo{chapter}{G. D. Mahan}, p.~\bibinfo{pages}{81}.

\bibitem[{\citenamefont{Jaccard et~al.}(1998)\citenamefont{Jaccard, Vargoz,
  Alami-Yadri, and Wilhelm}}]{jaccard.98}
\bibinfo{author}{\bibfnamefont{D.}~\bibnamefont{Jaccard}},
  \bibinfo{author}{\bibfnamefont{E.}~\bibnamefont{Vargoz}},
  \bibinfo{author}{\bibfnamefont{K.}~\bibnamefont{Alami-Yadri}},
  \bibnamefont{and} \bibinfo{author}{\bibfnamefont{H.}~\bibnamefont{Wilhelm}},
  \bibinfo{journal}{Rev. High Pressure Sci. Technol.}
  \textbf{\bibinfo{volume}{7}}, \bibinfo{pages}{412} (\bibinfo{year}{1998}).

\bibitem[{\citenamefont{Holems et~al.}(2004)\citenamefont{Holems, Jaccard, and
  Miyake}}]{holmes.04}
\bibinfo{author}{\bibfnamefont{A.~T.} \bibnamefont{Holems}},
  \bibinfo{author}{\bibfnamefont{D.}~\bibnamefont{Jaccard}}, \bibnamefont{and}
  \bibinfo{author}{\bibfnamefont{K.}~\bibnamefont{Miyake}},
  \bibinfo{journal}{Phys. Rev. B} \textbf{\bibinfo{volume}{69}},
  \bibinfo{pages}{024508} (\bibinfo{year}{2004}).

\bibitem[{\citenamefont{Wilhelm and Jaccard}(2002)}]{wilhelm.02}
\bibinfo{author}{\bibfnamefont{H.}~\bibnamefont{Wilhelm}} \bibnamefont{and}
  \bibinfo{author}{\bibfnamefont{D.}~\bibnamefont{Jaccard}},
  \bibinfo{journal}{Phys. Rev.} \textbf{\bibinfo{volume}{66}},
  \bibinfo{pages}{064428} (\bibinfo{year}{2002}).

\bibitem[{\citenamefont{Wilhelm et~al.}(2005)\citenamefont{Wilhelm, D.Jaccard,
  Zlatic, Monnier, Delley, and Coqblin}}]{Wilhelm.05}
\bibinfo{author}{\bibfnamefont{H.}~\bibnamefont{Wilhelm}},
  \bibinfo{author}{\bibnamefont{D.Jaccard}},
  \bibinfo{author}{\bibfnamefont{V.}~\bibnamefont{Zlatic}},
  \bibinfo{author}{\bibfnamefont{R.}~\bibnamefont{Monnier}},
  \bibinfo{author}{\bibfnamefont{B.}~\bibnamefont{Delley}}, \bibnamefont{and}
  \bibinfo{author}{\bibfnamefont{B.}~\bibnamefont{Coqblin}},
  \bibinfo{journal}{J. Phys: Cond. Matter} \textbf{\bibinfo{volume}{17}},
  \bibinfo{pages}{S830} (\bibinfo{year}{2005}).

\bibitem[{\citenamefont{Mahan}(1981)}]{mahan.81}
\bibinfo{author}{\bibfnamefont{G.~D.} \bibnamefont{Mahan}},
  \emph{\bibinfo{title}{Many-Particle Physics}} (\bibinfo{publisher}{Plenum},
  \bibinfo{address}{New York}, \bibinfo{year}{1981}).

\bibitem[{\citenamefont{Vidhyadhiraja and Logan}(2004)}]{logan.04}
\bibinfo{author}{\bibfnamefont{N.}~\bibnamefont{Vidhyadhiraja}}
  \bibnamefont{and} \bibinfo{author}{\bibfnamefont{D.}~\bibnamefont{Logan}},
  \bibinfo{journal}{European Phys. Jour. B} \textbf{\bibinfo{volume}{39}},
  \bibinfo{pages}{313} (\bibinfo{year}{2004}).

\bibitem[{\citenamefont{Logan and Vidhyadhiraja}(2005)}]{logan.05}
\bibinfo{author}{\bibfnamefont{D.~E.} \bibnamefont{Logan}} \bibnamefont{and}
  \bibinfo{author}{\bibfnamefont{N.~S.} \bibnamefont{Vidhyadhiraja}},
  \bibinfo{journal}{J. Phys: Cond. Matter} \textbf{\bibinfo{volume}{17}},
  \bibinfo{pages}{2935} (\bibinfo{year}{2005}).

\bibitem[{\citenamefont{Kernavanois et~al.}(2005)\citenamefont{Kernavanois,
  Raymond, Ressouche, Grenier, Flouquet, and Lejay}}]{kernavanois.2005}
\bibinfo{author}{\bibfnamefont{N.}~\bibnamefont{Kernavanois}},
  \bibinfo{author}{\bibfnamefont{S.}~\bibnamefont{Raymond}},
  \bibinfo{author}{\bibfnamefont{E.}~\bibnamefont{Ressouche}},
  \bibinfo{author}{\bibfnamefont{B.}~\bibnamefont{Grenier}},
  \bibinfo{author}{\bibfnamefont{J.}~\bibnamefont{Flouquet}}, \bibnamefont{and}
  \bibinfo{author}{\bibfnamefont{P.}~\bibnamefont{Lejay}},
  \bibinfo{journal}{Physical Review B} \textbf{\bibinfo{volume}{71}},
  \bibinfo{pages}{064404} (\bibinfo{year}{2005}).

\bibitem[{\citenamefont{Miyake and Maebashi}(2002)}]{miyake.02}
\bibinfo{author}{\bibfnamefont{K.}~\bibnamefont{Miyake}} \bibnamefont{and}
  \bibinfo{author}{\bibfnamefont{H.}~\bibnamefont{Maebashi}},
  \bibinfo{journal}{J. Phys. Soc. Japan} \textbf{\bibinfo{volume}{71}},
  \bibinfo{pages}{1007} (\bibinfo{year}{2002}).

\bibitem[{\citenamefont{Zlati\'c and Monnier}(2005)}]{zlatic.05}
\bibinfo{author}{\bibfnamefont{V.}~\bibnamefont{Zlati\'c}} \bibnamefont{and}
  \bibinfo{author}{\bibfnamefont{R.}~\bibnamefont{Monnier}},
  \bibinfo{journal}{Phys. Rev. B} \textbf{\bibinfo{volume}{71}},
  \bibinfo{pages}{165109} (\bibinfo{year}{2005}).

\bibitem[{\citenamefont{Alami-Yadri et~al.}(1999)\citenamefont{Alami-Yadri,
  Jaccard, and Andreica}}]{alami-yadri-99}
\bibinfo{author}{\bibfnamefont{K.}~\bibnamefont{Alami-Yadri}},
  \bibinfo{author}{\bibfnamefont{D.}~\bibnamefont{Jaccard}}, \bibnamefont{and}
  \bibinfo{author}{\bibfnamefont{D.}~\bibnamefont{Andreica}},
  \bibinfo{journal}{J. Low. Temp.Phys} \textbf{\bibinfo{volume}{114}},
  \bibinfo{pages}{135} (\bibinfo{year}{1999}).

\bibitem[{\citenamefont{Burdin and Zlati\'c}(2008)}]{burdin.08}
\bibinfo{author}{\bibfnamefont{S.}~\bibnamefont{Burdin}} \bibnamefont{and}
  \bibinfo{author}{\bibfnamefont{V.}~\bibnamefont{Zlati\'c}},
  \bibinfo{journal}{unpublished}  (\bibinfo{year}{2008}).

\bibitem[{\citenamefont{Khurana}(1990)}]{khurana.1990}
\bibinfo{author}{\bibfnamefont{A.}~\bibnamefont{Khurana}},
  \bibinfo{journal}{Phys. Rev. Lett.} \textbf{\bibinfo{volume}{64}},
  \bibinfo{pages}{1990} (\bibinfo{year}{1990}).

\bibitem[{\citenamefont{Zlati\'c and Horvati\'c}(1990)}]{zlatic-horvatic.1990}
\bibinfo{author}{\bibfnamefont{V.}~\bibnamefont{Zlati\'c}} \bibnamefont{and}
  \bibinfo{author}{\bibfnamefont{B.}~\bibnamefont{Horvati\'c}},
  \bibinfo{journal}{Solid State Commun.} \textbf{\bibinfo{volume}{75}},
  \bibinfo{pages}{263} (\bibinfo{year}{1990}).

\bibitem[{\citenamefont{Freericks and Zlati\ifmmode~\acute{c}\else
  \'{c}\fi{}}(2001)}]{freericks.01}
\bibinfo{author}{\bibfnamefont{J.~K.} \bibnamefont{Freericks}}
  \bibnamefont{and}
  \bibinfo{author}{\bibfnamefont{V.}~\bibnamefont{Zlati\ifmmode~\acute{c}\else
  \'{c}\fi{}}}, \bibinfo{journal}{Phys. Rev. B} \textbf{\bibinfo{volume}{64}},
  \bibinfo{pages}{245118} (\bibinfo{year}{2001}).

\bibitem[{\citenamefont{Horvati\ifmmode~\acute{c}\else \'{c}\fi{} and
  Zlati\ifmmode~\acute{c}\else \'{c}\fi{}}(1984)}]{horvatic.84}
\bibinfo{author}{\bibfnamefont{B.}~\bibnamefont{Horvati\ifmmode~\acute{c}\else
  \'{c}\fi{}}} \bibnamefont{and}
  \bibinfo{author}{\bibfnamefont{V.}~\bibnamefont{Zlati\ifmmode~\acute{c}\else
  \'{c}\fi{}}}, \bibinfo{journal}{Phys. Rev. B} \textbf{\bibinfo{volume}{30}},
  \bibinfo{pages}{6717} (\bibinfo{year}{1984}).

\end{thebibliography}

\hyphenation{Post-Script Sprin-ger}

\end{document}